\documentclass[aps,prd,eqsecnum,
epsf,nofootinbib]{revtex4}
\usepackage{amsfonts,latexsym,eucal,amsmath,amssymb,times,mathrsfs}
\usepackage[dvips]{graphicx}
\usepackage[usenames]{color}

\newcommand{\bm}[1]{\hbox{\boldmath{$#1$}}}
\newcommand{\sbm}[1]{\hbox{\boldmath{\scriptsize$#1$}}}
\newcommand{\Mp}{M_{\rm pl}}
\newcommand{\dd}{{\rm d}}
\newcommand{\gR}{{^g\!R}}
\newcommand{\sR}{{^s\!R}}
\newcommand{\sH}{{s}}
\newcommand{\Heisenberg}{_{\!\!{}_{H}}}
\newcommand{\HeisenbergH}{_{\!\!{}_{H\,H}\!\!}}
\newcommand{\gz}{{^g\!\zeta}}

\newcommand{\Gbz}{{^g\!\bar{\zeta}}}

\newcommand{\cv}{\{\zeta,\pi\}}
\newcommand{\cvt}{\{\tilde{\zeta},\tilde{\pi}\}}
\newcommand{\gauge}{{\em gauge} }
\newcommand{\cO}{{\cal O}}
\definecolor{red  }{rgb}{1,0,0}

\begin{document}

\thispagestyle{empty}


\title{Strong restriction on inflationary vacua from the local \gauge
invariance II:\\
Infrared regularity and absence of the secular growth in Euclidean vacuum
}
\date{\today}
\author{Takahiro Tanaka$^{1}$}
\email{tanaka_at_yukawa.kyoto-u.ac.jp}
\author{Yuko Urakawa$^{2}$}
\email{yurakawa_at_ffn.ub.es}
\affiliation{\,\\ \,\\
$^{1}$ Yukawa Institute for Theoretical Physics, Kyoto university,
  Kyoto, 606-8502, Japan\\
$^{2}$ Departament de F{\'\i}sica Fonamental i Institut de Ci{\`e}ncies del Cosmos, 
Universitat de Barcelona,
Mart{\'\i}\ i Franqu{\`e}s 1, 08028 Barcelona, Spain}

\preprint{200*-**-**, WU-AP/***/**, hep-th/*******}


\begin{abstract}
We investigate the initial state of the inflationary universe.
In our recent publications, we showed that requesting the
 \gauge invariance in the local observable universe to the initial state
  guarantees the
 infrared (IR) regularity of loop corrections in a general single clock
 inflation. Following this study, in this paper, we show that choosing
 the Euclidean vacuum ensures the \gauge
 invariance in the local universe and hence the IR regularity of loop
 corrections. It has been suggested that loop corrections to
 inflationary perturbations may yield the secular growth, which can
 lead to the break down of the perturbative analysis in an extremely long
 term inflation. The absence of the secular growth has been claimed by
 picking up only the IR contributions, which we think is 
 incomplete because the non-IR modes which are comparable to or smaller
 than the Hubble scale potentially can contribute to the secular
 growth. We prove the absence of the secular growth without neglecting
 these non-IR modes to a certain order in the perturbative expansion. 
 We also discuss how the regularity of the $n$-point functions for the genuinely \gauge invariant variable
 constrains the initial states of the inflationary universe.
 These results apply in a fully general single
 field model of inflation.     
\end{abstract}


\maketitle
\section{Introduction}  \label{Sec:Intro}
\subsection{Motivation and the current status of IR issues}
{\bf Initial states of the observable universe.~} 
How our universe began? This is one of the biggest question in
cosmology. The observation of the cosmic microwave background 
tells us about the cosmological perturbation
at the last scattering, which realistic
scenarios of the early universe should explain. If inflation took place
preceding the big-bang nucleosynthesis, the quantum 
fluctuation of the inflaton can generate the seed of 
the cosmological fluctuation which is consistent with the scale-invariant 
spectrum at large scales. Therefore, in the context of 
the inflationary scenario, which is currently the most successful 
scenario of the early universe, the period of inflation is 
the earliest part of our observable universe. In this
series of papers, we pursue the question; ``What can we claim about 
the initial quantum state of the observable universe if we require 
that the theoretical prediction should be stable against the infrared
(IR) loop contribution?'' 
The adiabatic vacuum is widely accepted to be the most natural vacuum
at least for a free field theory,
since it mimics the vacuum of the flat spacetime in
the ultraviolet (UV) limit. However, in a number of publications~\cite{Boyanovsky:2004gq, Boyanovsky:2004ph, Boyanovsky:2005sh,
Boyanovsky:2005px, Tsamis:1996qm, Tsamis:1996qq, Onemli:2002hr, Brunier:2004sb, Prokopec:2007ak,
Sloth:2006az, Sloth:2006nu, Seery:2007we, Seery:2007wf, Urakawa:2008rb,
Adshead:2008gk, Cogollo:2008bi, Rodriguez:2008hy, Seery:2009hs,
Gao:2009fx, Bartolo:2010bu, Seery:2010kh,
Kahya:2010xh, RB}, it has
been suggested that the adiabatic vacuum may not be stable against the
IR contributions in the presence of non-linear interactions.

{\bf Non-locality of the action and IR divergence problem.}
When we assume that the free field has the scale invariant spectrum in
the IR limit, a naive consideration can easily lead to the IR divergence
due to loop 
corrections. Here we illustrate how the IR divergence can appear from
the loop corrections of the curvature perturbation in single 
field models. Choosing
the time slicing on which the inflaton field is homogeneous, 
we can express the action in terms of the unique dynamical
degrees of freedom $\zeta$, the curvature perturbation, 
and the Lagrange multipliers $N$ and $N_i$, the lapse function
and the shift vector. The Hamiltonian and the momentum constraint 
equations relate the dynamical variable $\zeta$ 
to the multipliers $N$ and $N_i$. As is
explicitly shown in various papers, for instance in
Ref.~\cite{Maldacena2002, IRgauge_L, IRgauge}, these constraint equations 
are 
elliptic-type equations, and schematically written as
\begin{align}
 & \partial^2 N = f[\zeta]\,, \qquad  \partial^2 N_i = f_i[\zeta]\,,  \label{Eq:C}
\end{align}
where $\partial^2$ denotes the spatial Laplacian. By requesting the
regularity at the spatial infinity, the boundary conditions of these
elliptic-type equations are uniquely fixed. 
Substituting the expressions of $N$ and $N_i$ into the action, 
we obtain
\begin{align}
 & S= \int \dd^4 x\, {\cal L} [\zeta,\, N,\, N_i]
 = \int \dd^4 x\, {\cal L}[\zeta,\, \partial^{-2} f[\zeta],\,
 \partial^{-2} f_i[\zeta] ]\,, \label{Exp:Sz}
\end{align} 
and hence the evolution of $\zeta$ is described by the above 
{\it non-local} action. Here the inverse Laplacian $\partial^{-2}$ 
is usually supposed to be defined as multiplying the inverse of the 
eigenvalue of the Laplacian operator 
by using the harmonic decomposition. When we evaluate the loop corrections
to the $n$-point functions expanding them in terms of the interaction
picture field $\zeta_I$, we need to evaluate the expectation values such as 
\begin{align}
 & \langle \zeta_I^2 \rangle \,, \qquad \langle \zeta_I \partial^{-2} \zeta_I
 \rangle \,,\qquad \cdots \,.  \label{ansatz}
\end{align}
Inserting the scale invariant spectrum into 
$\langle \zeta_I^2 \rangle$ 
leads to the logarithmic divergence as
$\langle \zeta_I^2 \rangle \propto \int \dd^3 \bm{k}/k^3$. The second
expression of Eq.~(\ref{ansatz}),
which may arise as a consequence of the operation of $\partial^{-2}$, 
is more singular as
$\langle \zeta_I \partial^{-2} \zeta_I \rangle \propto \int \dd^3 \bm{k}/k^5$,
which diverges quadratically. 
The presence of non-local interactions enhances the long range 
correlations, and hence the singular behaviour in the IR. 
When we introduce the
IR cutoff, say at the Hubble scale at a particular time $t_0$, 
the variance $\langle \zeta_I^2 \rangle$ shows 
the logarithmic secular growth as
$\langle \zeta_I^2 \rangle \propto \int^{aH}_{a_0 H_0} \dd k/k \sim \log a/a_0$
where $a_0$ and $H_0$, respectively, 
denote the scale factor and the Hubble scale at
$t=t_0$. If the IR divergence exists, the loop
corrections, which are suppressed by an extra power of the amplitude of the
power spectrum $(H/\Mp)^2$, may dominate
in case inflation continues sufficiently
long, leading to the break down of perturbation.

{\bf The dilatation symmetry as a necessary ingredient for IR regularity.}
The regularization of the IR contributions has been discussed in a
number of publications~\cite{IRsingle, IRmulti, IRgauge_L, IRgauge,
IRgauge_multi, SRV1, BGHNT10, GHT11, GS10, GS11, GS1109, PSZ, SZ1203, SZ1210}. The important
aspect in discussing the long wavelength mode of $\zeta$ is the
dilatation symmetry of the system. As is expected from the fact that the
spatial metric is given in the form $a^2 e^{2\zeta}  \dd \bm{x}^2$,
a constant shift of the dynamical variable $\zeta$ can be absorbed by 
the overall rescaling of the spatial coordinates. 
Hence, the action for $\zeta$ preserves the dilatation symmetry:
\begin{align}
 & x^i \to e^{-s} x^i\,, \qquad  \zeta(t,\, \bm{x}) \to
 \zeta(t,\, e^{-s} \bm{x}) - s\,,
\end{align}
where $s$ is a constant parameter. (There are a number of
literatures where this dilatation symmetry is addressed. See for
instance, Refs.~\cite{Creminelli:2012ed, Hinterbichler:2012nm} and the
references therein.) One may naively expect that
we can absorb the IR divergent contribution of $\zeta$ 
using this constant shift. 
As an example, we set the parameter $s$ to the spatial
average of the curvature perturbation within the Hubble patch at $t_0$,
$\bar{\zeta}(t_0)$, where the size of the Hubble patch in comoving
coordinates is given by $1/(a_0 H_0)$. Then, the logarithmically
divergent two-point function $\langle \zeta_I^2 \rangle$ seems to be
replaced with  
$
\langle (\zeta_I - \bar{\zeta}_I)^2 \rangle 
\propto \int^{aH}_{a_0 H_0} \dd k/k$\,, which is finite but still grows
logarithmically in time. 
One may think that if the system is described in such a way that the
symmetry under the time dependent dilatation transformation is manifest, 
setting $s(t)$ to the time dependent spatial average in the Hubble patch, 
the logarithmic growth of $\bar{\zeta}(t)$ might be eliminated. 
However, the reduced action written in terms of $\zeta$
(\ref{Exp:Sz}) does not preserve the invariance under the dilatation 
transformation with the time dependent parameter $s(t)$. 
For example, in the recent
literature~\cite{Hinterbichler:2012nm}, the
authors showed that when we consider the whole universe with the
infinite spatial volume, the dilatation transformation should be time
independent to preserve the action invariant. 
In addition, the two-point function with
$\partial^{-2}$ cannot be regularized by considering the dilatation 
symmetry alone. This quick
consideration tells us that the presence of the dilatation symmetry of
the system may play an important role in the regularization of the IR
contributions but is not sufficient to guarantee the IR 
regularity and the absence of the secular growth.

{\bf Residual \gauge degrees of freedom in the local universe.} 
A missing piece in the above discussion is to pay careful attention 
to what are the quantities we can actually observe. Since our 
observable region is a limited portion of the whole universe, 
the observable fluctuations must be composed of local quantities. 
Furthermore, as the information that we can access 
is limited to our observable region, 
there is no reason to request the regularity at the
spatial infinity in solving the elliptic constraint equations (\ref{Eq:C}). 
Then, there arise degrees of freedom in choosing the boundary
conditions of Eqs. (\ref{Eq:C}). The
degrees of freedom in solutions of $N$ and $N_i$ can be understood as 
the degrees of freedom in choosing coordinates. As we showed in
Refs.~\cite{IRgauge_L, IRgauge}, these residual coordinate
transformations are expressed in terms of homogeneous solutions 
to the Laplace equation as
\begin{align}
 &  x^i \to x^i - s(t) x^i - \sum_{m=1} {s^i}_{j_1 \cdots j_m}(t) x^{j_1}
 \cdots x^{j_m} + \cdots\,, \label{Exp:RC} 
\end{align}
where ${s^i}_{j_1 \cdots j_m}(t)$ are symmetric traceless tensors, which
satisfy 
$\delta^{j j'} {s^i}_{j_1 \cdots j \cdots j'  \cdots j_m}(t)=0$. 
Here, we abbreviated the non-linear terms in the coordinate
transformation. Note that this coordinate transformations include the
dilatation transformation with the time dependent function $s(t)$. 
Since the transformations in Eq.~(\ref{Exp:RC}) are nothing but 
coordinate transformations, the diffeomorphic invariant action 
$S=\int \dd^4 x {\cal L}[\zeta,\, N,\, N_i]$ should preserve the
symmetry under these transformations. Thus, when we consider only the local
observable region, which is a potion of the whole universe, we find an infinite
number of coordinate transformations which keep the action
invariant. Considering the dilatation transformation in the whole universe
is subtle in the sense that the transformation diverges at the spatial
infinity, even if the parameter $s$ is very small. By contrast,
restricted to the local
region, the magnitude of the coordinate transformations in Eq.~(\ref{Exp:RC})
is kept perturbatively small. In this paper, we refer to the local observable
(spacetime) region as ${\cal O}$. 
The size of the observable region on each time slicing is supposed to be of order  $1/a(t)H(t)$ 
at least in the far past since the past light cone asymptotes to 
that size.
We should note that, once we insert the expressions of $N$ and $N_i$
into the action to obtain the action for the curvature perturbation
$\zeta$, the symmetry under the residual coordinates transformation is
lost, because specific boundary conditions  are chosen for $N$ and $N_i$ in
fixing coordinates. To emphasize the distinction between the
coordinate transformations associated with the change of the boundary
conditions and the usual gauge transformation, which keeps
the action invariant, we 
denote the former by the \gauge transformation in the italic font.

{\bf Removing the residual \gauge degrees of freedom.}
One way to realize the invariance under the gauge transformation is
fixing the gauge conditions completely. The residual
\gauge degrees of freedom introduced above can be also 
removed by employing additional \gauge conditions, 
{\it i.e.,} by fixing the boundary
conditions of $N$ and $N_i$ at the boundary of the local 
region ${\cal O}$. Then, 
we naturally expect that 
the IR regularity may be explicitly shown by performing the quantization in this
local region, since the wavelengths that fit within this local
region ${\cal O}$ will be bounded by the size of the region. 
Although the quantization in the local region is an interesting approach, 
it is not so clear how to perform the quantization after removing the 
residual \gauge degrees of freedom. One of the difficulties is
that even the translation symmetry of 
the quantum state cannot be easily guaranteed in the local system, 
since it is manifestly broken by introducing the 
boundary condition at a finite distance. 
(See also the discussion in Ref.~\cite{SRV1}). 

As an alternative way, in Ref.~\cite{IRsingle}, we first set the initial
state considering the whole universe, and then we performed the 
residual \gauge transformation (\ref{Exp:RC}) to fix the coordinates
so that the IR contributions are absorbed. 
Through the transformation with 
$s(t)=\bar\zeta(t)$, 
the curvature perturbation is transmitted as
\begin{align}
 & \zeta(t,\, \bm{x}) \to  \zeta(t,\,
 e^{-\bar{\zeta}(t)} \bm{x}) - \bar{\zeta}(t) =  \zeta(t,\, \bm{x}) -
 \bar{\zeta}(t) + {\cO}(\zeta^2) \,. \label{Exp:RC2}
\end{align}
Here, $\zeta(x)$ is the original curvature perturbation defined in the
whole universe and 
its spatial average over the whole universe is set to 0 
as in the conventional cosmological perturbation theory.
By contrast, 
$\zeta(t,\, e^{-\bar{\zeta}(t)} \bm{x}) -\bar{\zeta}(t)$ is the 
curvature perturbation relevant to the local universe, and its spatial
average over the local region $\Sigma_t \cap {\cal O}$ is set to $0$,
where $\Sigma_t$ is a time constant surface.
In Ref.~\cite{IRsingle} we considered the fluctuation of the inflaton,
using the flat gauge, but the same discussion follows also for the curvature
perturbation $\zeta$. 
In the recent publication by Senatore and Zaldarriaga~\cite{SZ1210}, the same
degrees of freedom in choosing coordinates are used in a slightly
different way to absorb the IR divergent contributions. 
If the non-linear terms in the 
residual \gauge transformation at the initial 
time~(\ref{Exp:RC2}) did not yield IR divergent contributions, 
the discussion in Ref.~\cite{IRsingle} would have proved the absence of 
IR divergence in general. 
What was shown there is that 
once the field operator after the residual \gauge transformation 
is guaranteed to be regular at the initial time, its succeeding 
evolution does not produce IR divergence. 
The heart of the proof is that 
$\zeta_I(x)$ is replaced with $\zeta_I(x) -
\bar{\zeta}_I(t)$ in the expansion of the composite operators  
in terms of the interaction picture field, 
after the residual \gauge transformation, and hence 
the IR contributions from $\zeta_I(x)$ are always canceled by those
from $\bar{\zeta}_I(t)$.
However, the non-linear part of the transformation at the initial time 
contains $\bar{\zeta}(t)$ whose IR contributions logarithmically
diverge. 
The lesson is that it is not straightforward to reformulate the
way of quantization so that the IR divergent contributions therein 
are all absorbed by
the residual \gauge transformation. (The absorption of the IR modes
of the curvature perturbation was intended in other frameworks such as
$\delta N$ formalism~\cite{BGHNT10, GHT11} and the semi-classical approach~\cite{GS10}. )

{\bf The secular growth.}
The appearance of IR divergence due to the
residual \gauge transformation mentioned above 
might be evaded 
by sending the initial time to the past infinity. 
This is because the size of the local region
$\Sigma_t \cap {\cal O}$ in comoving coordinates becomes infinitely large in this
limit, making the discrepancy between the average in the local region and 
that in the global universe smaller and smaller. Then 
it might be effectively unnecessary to perform the residual \gauge
transformation at the initial time, although this statement is not very
rigorous.  
We should note that when we send the initial time to the past infinity, 
it is too naive to neglect the non-IR modes which
are comparable to or shorter than the Hubble length scale, 
because all the modes were much shorter than the Hubble length
scale in the distant past. 
This makes the issue regarding the secular growth much more complicated. 
For instance, once we include the contributions from the non-IR modes, 
we cannot 
use the conservation of $\zeta_{\sbm{k}}$ in the limit $k/aH \ll 1$, 
where $\bm{k}$ is the comoving wavenumber of the external leg, 
relying on the long wavelength approximation such as $\delta N$ formalism. 
Here, in a simple example,
we show that vertex integrations can yield the apparent secular growth through 
the non-linear contributions from the modes at around the Hubble scale. 
Even if the vertex is confined in the region ${\cal O}$, 
the integration region of each vertex is still infinite in the time direction as
$
\int \dd t \dd^3 \bm{x} a^3 (\cdots) \simeq \int \dd (\ln a)/H^4
(\cdots)
$,
which may cause the secular growth. 
Roughly speaking, 
the integrand $(\cdots)$ will be written in terms of the dimensionless
time dependent slow roll parameters and the wavenumber
of the fields in this vertex $k_m/aH$ normalized by the Hubble scale. 
If we focus on the non-linear interaction composed of the modes with 
$k_m/aH$ of order unity, the
integrand $(\cdots)$ are expressed only in terms of the parameters which
are supposed to change very slowly in time and then the contribution
from the interaction vertex seems to yield the logarithmic growth. 
This is another origin of the secular growth, which should be
distinguished from the one inherited from the IR behavior of  
$\langle (\zeta_I)^2 \rangle$. 
Of course the above argument is too native, but it shows that 
the absence of the secular growth from the vertex integration 
is rather subtle, requiring more careful treatment about the modes 
around the Hubble scale. 
Because of this subtlety, introducing the UV cutoff at the length 
scale longer or equal to the Hubble length scale by hand makes 
the discussion incomplete. 
In fact, if it were allowed to simply neglect the short 
wavelength modes, the discussion in Ref.~\cite{IRsingle} with the
initial time $t_i$ sent to $-\infty$ would 
have given a rough proof of the absence of IR divergence 
without any limitation to the quantum state by sending the 
initial time to the past infinity, which contradicts 
our current claim that the quantum state is restricted in order 
to avoid IR divergence. Recently,
the absence of the secular growth was claimed relying on the conservation of
the curvature perturbation in Refs.~\cite{PSZ, SZ1210}, but the
aspects mentioned above were not discussed. 
In addition, even if the conservation of 
$\zeta_{\sbm{k}}$ in the limit $k/aH \ll 1$ is proved, 
the logarithmic enhancement in the form $(k/aH)^2 
\ln (k/a_i H_i)$ may
give rise, where $a_i$ and $H_i$ are the scale factor and the
Hubble parameter at the initial time. The factor 
$\ln(k/a_i H_i)$ can become large to overcome the suppression by $(k/aH)$ 
when we send the initial time to the past infinity.

\subsection{Summary of upcoming results}
{\bf Short summary of the results. } In this subsection, we summarize what we will show in this paper. 
Taking account of the current status of IR issues mentioned above,  
we will establish the following three statements in this paper: 
\begin{enumerate}
 \item There is an alternative equivalent Hamiltonian that describes the quantum
       dynamics of our interest and whose interaction part is 
       solely composed of the IR irrelevant operators(, which mean the
       field operators associated with the operations that manifestly
       suppress the IR contribution such as $\partial_i/aH$ and $\partial_t/H$).
 \item The Euclidean vacuum state, which is specified by the regularity
       when the time coordinates in the $n$-point functions are
       analytically continued to the imaginary in the complex plane, is physically the same 
       both in the alternative description mentioned
       in item 1, and in the original description. 
 \item The $n$-point functions in the Euclidean vacuum state 
       respect the spatial translation invariance and are regular
       in the IR. The secular growth is absent, even if we include the vertices 
       with non-IR modes, as long as very high order of loop corrections
       are not concerned.  
\end{enumerate}
Below we add a little more detailed 
explanations about the above three items. 

{\bf {\em Gauge} issue.} 
In this paper, the quantization and fixing the initial quantum state as
a starting point of our discussion is performed in the original system which describes the whole
universe, where the residual \gauge degrees of freedom are left unfixed. 
Then, following Refs.~\cite{IRgauge_L, IRgauge, IRgauge_multi, SRV1,IRNG}, we
introduce a field operator which preserves the invariance under any spatial 
coordinates transformations, including residual
\gauge transformations. We refer to such an operator as a genuine
\gauge invariant operator. As a representative, we consider a genuine 
\gauge invariant curvature perturbation, $\gR$. 
As long as the expectation values of such genuine gauge invariant operators are  
concerned, we can perform the residual \gauge transformation
without affecting the results of computations.
We will show that, using this residual \gauge transformation,  
the boundary conditions of 
the non-local operator $\partial^{-2}$ in the action 
can be modified to be regular in the IR. 

{\bf Requirement of the {\gauge} invariance in quantum state.} 
To calculate the $n$-point functions which preserve
the invariance under the residual \gauge transformations, 
the initial state should be also specified in a genuinely \gauge
invariant manner. 
However, when we perform the quantization considering the
whole universe, preserving the residual \gauge 
invariance becomes obscure, because these residual \gauge degrees of freedom
are not present as long as we deal with the whole universe. 
In our previous paper~\cite{SRV1}, we discovered 
a correspondence between the IR regularity and the invariance
under the residual \gauge transformations, which will provide 
an important clue to the guiding
principle in choosing the genuinely \gauge invariant initial state. 
To discuss this point, aside from the original canonical variables
$\zeta(x)$ and its conjugate momentum 
$\pi(x)$, whose evolution is governed by the action (\ref{Exp:Sz}), we
introduced another set of the canonical variables corresponding to the 
description in the coordinates
shifted by a constant dilatation transformation:
\begin{align}
 & \tilde{\zeta}(x):= \zeta(t,\, e^{-s} \bm{x})\,, \qquad
 \tilde{\pi}(x)\,,
\end{align}
where $s$ is a time independent c-number and 
$\tilde{\pi}(x)$ is the conjugate momentum of $\tilde{\zeta}(x)$.
In Ref.~\cite{SRV1}, 
we showed that requesting the equivalence between the two quantum 
systems described by $\cv$ and $\cvt$ guarantees the 
IR regularity of loop corrections. Here,  
the equivalence of two quantum systems means that the same 
iteration scheme (or formally the same initial condition of the 
interacting system) gives physically the same quantum state in both 
systems related to each other by the dilatation transformation. 
Namely, all the expectation values evaluated in both systems are 
equivalent if we take into account how they transform under 
dilatation transformation. Requesting this equivalence will be thought of as the invariance of 
the initial state under the dilatation transformation. 
In Ref.~\cite{SRV1}, we employed the 
iteration scheme in which the interaction is turned on at a
finite past. 
Then, it turned out that the IR regularity/\gauge invariance condition
cannot be consistently imposed. In the present paper we will set the initial quantum
state at the infinite past. We will show that the above transformation
can be extended to allow a time dependence of the parameter $s$. As we
described in the previous section, this extension plays a crucial role
in discussing the absence of the secular growth.  

{\bf The Euclidean vacuum.} 
The second and third items are related with each other, once we 
establish the correspondence between the \gauge invariance and the 
IR regularity. We will show that the two quantum systems described by $\cv$ and $\cvt$ are 
equivalent if we choose the Euclidean vacuum, which is defined by  
requesting the regularity of the $n$-point functions at the
distant past with the time path rotated toward the complex plane. 
To be more specific, as the second item, we
will show that the $n$-point functions for $\zeta(x)$ calculated by the
canonical variables $\cv$ with the boundary condition of the
Euclidean vacuum agrees
with the $n$-point functions for $\tilde{\zeta}(t, e^{s(t)} \bm{x})$ calculated by the
canonical variables $\cvt$ under formally the 
same boundary condition, {\it i.e.},
\begin{align}
 & \langle\, \zeta(t,\, \bm{x}_1) \zeta(t,\, \bm{x}_2) \cdots \zeta(t,\,
 \bm{x}_n)\, \rangle_{\cv} =  \langle\, \tilde\zeta(t,\, e^{s(t)} \bm{x}_1) \tilde\zeta(t,\,
  e^{s(t)} \bm{x}_2) \cdots \tilde\zeta(t,\,  e^{s(t)} \bm{x}_n)\,
 \rangle_{\cvt}\,.  \label{Exp:npI}
\end{align}
Combined with the previously mentioned
technique to deal with the \gauge issue, we
will show that when we choose the Euclidean vacuum, the Hamiltonian density for $\cvt$, 
can be expressed only in terms of the IR irrelevant operators.

{\bf The IR regularity and the absence of the secular growth.}
As for the third item, we evaluate the $n$-point function of the genuinely
\gauge invariant operator. Performing the quantization in the canonical system of
$\cvt$, we will show that the IR contributions do
not diverge and that the secular growth is suppressed. 
We carefully investigate 
the contributions from the modes which are
comparable to or less than the Hubble scale, {\it i.e.}, 
$k \gtrsim aH$, without employing the asymptotic
expansion with respect to $k/aH$. 
As is stressed at the end of the preceding subsection, 
this point is one of the necessary ingredients to show the absence of the secular
growth. One may naively expect that the UV modes with 
$k/aH \gtrsim 1$ will not effectively contribute to the vertex integration because of the
oscillatory behaviour. A more careful consideration tells us that this
naive expectation is not necessarily correct. In general, vertex
integrations become a mixture of the positive and negative
frequency mode functions, which yields 
the phase in the UV
limit $e^{i \eta (k_1 - k_2 + k_3 - \cdots)}$ where $\eta$ represents the
conformal time which runs from $-\infty$ to 0. 
Then, the phase does not necessarily exhibit the rapid
oscillation even for the modes with $k_m/aH \simeq - k_m \eta \gtrsim 1$,
where $m = 1,\, 2,\, \cdots$, which can be a cause of secular growth. 
Intriguingly, choosing the Euclidean
vacuum plays a crucial role not only in the IR limit but also in the UV
limit. 
One can show that there is no mixing between the positive and the negative
frequency modes, if we choose the Euclidean vacuum. 
Therefore, secular growth is evaded in this case. 

{\bf The outline of the paper.}
The outline of this paper is as follows. In Sec.~\ref{Sec:preparation},
we will briefly review the way to construct the genuinely \gauge
invariant operator $\gR$, following Refs.~\cite{IRgauge_L,
IRgauge}. Then, we will introduce the canonical variables 
$\cvt$ and will derive the Hamiltonian for these
variables. In Sec.~\ref{Sec:Inf}, we will discuss the items 1 and
2 that we mentioned above. In Sec.~\ref{SSec:Euclidean}, we will
describe the boundary conditions of the Euclidean vacuum and will prove 
Eq.~(\ref{Exp:npI}), which implies that the boundary
conditions of the Euclidean vacuum select the same ground state both in
$\cv$ and $\cvt$. 
In Sec.~\ref{SSec:np} and Sec.~\ref{RLR}, we will formulate the canonical quantization in terms
of $\cvt$ and will show that the interacting vertices
for these canonical variables consist only of the IR irrelevant
operators. Particularly in Sec.~\ref{RLR}, we will show that using the
residual \gauge degrees of freedom, the non-local operator
$\partial^{-2}$ can be made IR regular. In Sec.~\ref{Sec:IRregularity}, we will discuss the item 3.
In Sec.~\ref{sec:iepsilon}, we will show that the boundary
condition of the Euclidean vacuum leads to the so-called $i\epsilon$
prescription in a perturbative expansion.
In Sec.~\ref{SSec:Wightman}, we will calculate the
Wightman propagator, by which the $n$-point functions are 
expanded. Then, in Sec.~\ref{SSec:regularity}, we explicitly evaluate
$n$-point functions to investigate the IR regularity and the
secular growth. In Sec.~\ref{Sec:C}, as concluding remarks, we
discuss another possibility of the initial state which satisfies 
the IR regularity/\gauge invariance conditions. We 
will also mention the related papers to clarify what is new in this paper.

{\bf The advantage of the in-in formalism.}
In our previous publications~\cite{IRsingle, IRmulti, SRV1}, in calculating $n$-point
functions, we used the retarded Green function to solve the non-linear 
Heisenberg equation. This is because we thought that using the retarded
Green function, whose Fourier mode is regular in the IR limit,    
makes the proof of the IR regularity transparent. 
However, the perturbative expansion using the
retarded Green function is not suitable for the present purpose, 
because the positive and negative
frequency modes are mixed in the vertex integrations 
once the retarded Green function is used. Therefore, the
boundary conditions of the Euclidean vacuum does not guarantee the
convergence of the time integrations for all the vertices. By contrast, when we
calculate the $n$-point functions in the in-in formalism, all vertex
integrals can be made manifestly convergent by adopting the boundary 
conditions of the
Euclidean vacuum (see Sec.~\ref{SSec:Euclidean}). Since the $n$-point
functions obtained from the solution written in terms of the retarded Green
function agree with those obtained in the in-in formalism, the vertices
which do not converge should vanish in the final result of the
$n$-point functions. However, the cancellation is obscured in an
explicit perturbative expansion. Therefore, in this paper,
we calculate the $n$-point function totally based on
the in-in formalism, without using the retarded Green function. 

\section{Constructing the \gauge invariant quantity}
\label{Sec:preparation}
In this paper, as an explicit model of inflation, we consider a standard
single field inflation model whose action takes the form   
\begin{eqnarray}
 S = \frac{\Mp^2}{2} \int \sqrt{-g}~ [R - g^{\mu\nu}\phi_{,\mu} \phi_{,\nu} 
   - 2 V(\phi) ] \dd^4x~, \label{Exp:action} 
\end{eqnarray}
where $\Mp$ is the Planck mass and we set $\phi$ to a dimensionless scalar field,
dividing it by $\Mp$. However, as long as we consider a scalar field with
the second-order kinetic term, an extension proceeds in a
straightforward way. In
Sec.~\ref{SSec:Quantization}, we will construct the genuine \gauge
invariant operator corresponding to the spatial curvature of 
a $\phi$-constant surface. In 
Sec.~\ref{SSec:dilatation}, we will introduce the canonical system 
$\cvt$ whose Hamiltonian density is composed only
of the IR irrelevant operators. 

\subsection{{\em Gauge } invariant operator and quantization}
\label{SSec:Quantization}
We fix the time slicing by adopting the uniform field gauge
$\delta\phi=0$. Under the ADM metric decomposition, which is given by
\begin{eqnarray}
 \dd s^2 = - N^2 \dd t^2  + h_{ij} (\dd x^i + N^i \dd t) (\dd x^j + N^j
  \dd t)~, \label{Exp:ADMmetric}
\end{eqnarray}
we take the spatial metric $h_{ij}$ as
\begin{align}
 & h_{ij} = e^{2(\rho+\zeta)} \left[ e^{\delta \gamma} \right]_{ij}\,,
\end{align}
where $a:=e^\rho$ is the scale factor, $\zeta$ is the
so-called curvature perturbation and $\delta \gamma_{ij}$ is a traceless
tensor:
\begin{align}
 \delta {\gamma^i}_i=0\,.
\end{align}  
As spatial gauge conditions we impose 
the transverse conditions on $\delta \gamma_{ij}$:
\begin{align}
 & \partial_i \delta \gamma^i\!_j=0\, . 
\label{TTgauge}
\end{align}
Since the time slicing is fixed by the gauge condition $\delta \phi=0$,
there are remaining residual \gauge 
degrees of freedom only in choosing the spatial coordinates. 
In this paper, we neglect the vector and tensor perturbations. The
tensor perturbation, which is massless, can also contribute to
the IR divergence of loop corrections. We will address this issue in our
future publication. 

Following Refs.~\cite{IRgauge_L, IRgauge}, we 
construct a genuine \gauge invariant operator, which preserves the \gauge
invariance in the local observable universe. For the construction, we note that the scalar curvature $\sR$, which transforms as a scalar
quantity under spatial coordinate transformations, becomes genuinely
\gauge invariant, if 
we evaluate it in the geodesic normal coordinates on each time slice. 
The geodesic normal coordinates are 
introduced by solving the spatial
three-dimensional geodesic equation:  
\begin{eqnarray}
 \frac{\dd^2 x_{gl}^i}{\dd \lambda^2} +  {^s \Gamma^i}_{jk} \frac{\dd
  x_{gl}^j}{\dd \lambda} \frac{\dd x_{gl}^k}{\dd \lambda} =0~,
\label{GE}
\end{eqnarray}
where ${^s \Gamma^i}_{jk}$ is the Christoffel symbol with respect to 
the three dimensional spatial metric on a constant time hypersurface and
$\lambda$ is the affine parameter. 
Here we put the index $gl$ on the global coordinates, to reserve 
the simple notation $\bm{x}$ for the geodesic normal coordinates, which will be
mainly used in this paper. 
We consider the three-dimensional geodesics whose affine parameter ranges
from $\lambda=0$ to $1$ with the initial ``velocity'' given by
\begin{eqnarray}
 \frac{\dd x^i_{gl}(\bm{x},\lambda)}{\dd \lambda} \bigg\vert_{\lambda=0}= e^{-\zeta(\lambda=0)}
 \bm{x}^i\,. \label{IC}
\end{eqnarray}
A point $x^i$ in the geodesic normal coordinates is identified 
with the end point of the geodesic, $x_{gl}^i(\bm{x},\lambda=1)$ in the 
original coordinates. 
Using the geodesic normal coordinates $x^i$, we perturbatively 
expand $x_{gl}^i$ as 
$x_{gl}^i= x^i + \delta x^i(\bm{x})$. 
Then, we can construct a genuinely \gauge invariant
variable as 
\begin{align}
 {^g\!R}(t,\,\bm{x}) &:= \sR (t,\, x^i_{gl}(\bm{x})) = 
 \sR (t,\, x^i + \delta x^i (\bm{x})))\,, \label{Def:gR}
\end{align}
where $t$ denotes the cosmological time.

\subsection{Dilatation symmetry in the global universe}
\label{SSec:dilatation}
The focus of this subsection is on the dilatation transformation, 
shifting to the rescaled spatial coordinates: 
\begin{equation}
\tilde x^i :=e^{s(t)} x^i.
\label{cT}
\end{equation}  
Solving the Hamiltonian and momentum constraint equations, we can derive
the action that is expressed only in terms
of the curvature perturbation $\zeta(x)$, 
which is schematically written as 
\begin{align}
 S  &=  
\int \dd t\, \dd^3 \bm{x}\,  
   {\cal L} [\partial_t \zeta(x), \zeta(x)]\,,
\end{align}
Using the curvature perturbation $\zeta$ and the conjugate momentum defined by 
$\pi := \delta {\cal L}/\delta (\partial_t {\zeta})$, 
the Hamiltonian density is given by the Legendre transform as
\begin{align}
 & {\cal H}[\zeta(x), \pi(x)]
:=  \pi(x)
 \partial_t \zeta(x) -{\cal L}[\partial_t\zeta(x),\,\zeta(x)] 
\,. 
\end{align}

What is important here is only the fact that 
the curvature perturbation $\zeta$ appear in the action either with 
differentiation or in the form of the combination 
of the physical distance $e^{\rho+ \zeta}\dd \bm{x}$~\cite{SRV1}. 
In the new coordinates \eqref{cT}, 
the physical distance is written as 
$e^{\rho+ \tilde \zeta(t,\tilde{\sbm{x}})-s(t)} \dd \tilde{\bm{x}}$,
with the definition of a new variable 
\begin{equation}
 \tilde{\zeta}(t,\, \tilde{\bm{x}}) := 
   \zeta(t,\, \bm{x})~.
\label{deftildezeta}
\end{equation}
Thus, if the field 
$\zeta(x)$ is replaced with $\tilde\zeta(t,\,\tilde{\bm{x}})-s(t)$
under the change of the coordinates from $\bm{x}$ to $\tilde{\bm{x}}$, 
the action basically remains invariant. 
To express $\partial_t \zeta(x)$ in terms of the new variable
$\tilde\zeta$, we denote
the partial differentiation with the spatial coordinates $\bm{x}$ fixed as 
$({\partial}_t \tilde\zeta(t,\, \tilde{\bm{x}}))_{\sbm{x}}$. 
The subscript associated with the parentheses 
specifies the spatial coordinates that we fix in 
taking the partial differentiation. Then, we have 
\begin{align}
 & ({\partial}_t \tilde\zeta(t,\, \tilde{\bm{x}}))_{\sbm{x}}
 = \partial_t\zeta(x)  \,. 
\end{align}
For brevity, when the fixed spatial coordinates are identical to the ones in 
the argument of the variable, we simply use $\partial_t$. 
Then, we can establish an identity 
\begin{equation}
\int \dd t\, \dd^3 \bm{x}\,  
   {\cal L} [\partial_t \zeta(x), \zeta(x)]
=
\int \dd t\, \dd^3 \tilde{\bm{x}}\,  
   {\cal L} [(\partial_t \tilde\zeta(t, \tilde{\bm{x}}))_{\sbm{x}}, 
 \tilde\zeta(t, \tilde{\bm{x}})-s(t)]~.
\end{equation}
Recalling the relation between $\bm{x}$ and $\tilde{\bm{x}}$ (\ref{cT}), 
this equality also means the equality at the level 
of Lagrangian density, 
$e^{-3s(t)}{\cal L} [\partial_t \zeta(x), \zeta(x)]
={\cal L} [(\partial_t \tilde\zeta(t, \tilde{\bm{x}}))_{\sbm{x}},
 \tilde\zeta(t, \tilde{\bm{x}})-s(t)]$.

We introduce the canonical conjugate momentum corresponding to 
$\tilde\zeta(t,\tilde{\bm{x}})$ in the standard way as
\begin{align}
 \tilde{\pi}(t,\, \bm{x})
:= \frac{\partial {\cal L}[(\partial_t\tilde\zeta(t,\, \tilde{\bm{x}}))_{\sbm{x}}, \tilde\zeta(t,\, \tilde{\bm{x}})-s(t)]}{\partial 
(\partial_t \tilde\zeta(t,\, \tilde{\bm{x}}))}\,. 
\end{align}
Noticing the relation 
\begin{align}
 & \partial_t \tilde{\zeta}(t,\tilde{\bm{x}})= 
 (\partial_t \tilde{\zeta}(t,\tilde{\bm{x}}))_{\sbm{x}}
 - \dot{s}(t)\, \tilde{\bm{x}} \cdot
 \partial_{\tilde{\sbm{x}}} \tilde\zeta(t,\tilde{\bm{x}})\,,  \label{Exp:temp}
\end{align}
we have 
\begin{align}
 \tilde{\pi}(t,\, \tilde{\bm{x}})= 
\frac{\partial {\cal L}
[(\partial_t\tilde\zeta(t,\, \tilde{\bm{x}}))_{\sbm{x}}, \tilde\zeta(t,\, \tilde{\bm{x}})-s(t)]
}{\partial 
((\partial_t \tilde\zeta(t,\, \tilde{\bm{x}}))_{\sbm{x}})}
=e^{-3s(t)}
\frac{\partial {\cal L}[\partial_t 
\zeta(x), \zeta(x)]}{\partial 
(\partial_t \zeta(x))}
=e^{-3s(t)}\pi(x)
\,. 
\label{rel}
\end{align}
As is expected,
using the commutation relations for $\zeta$ and $\pi$ together
with Eqs.~(\ref{deftildezeta}) and (\ref{rel}), 
we can verify  
\begin{align}
 \left[ \tilde{\zeta}(t,\, \tilde{\bm x}),\, \tilde{\pi}(t,\, \tilde{\bm y})
 \right] = e^{-3s(t)} i \delta^{(3)}(({\bm x} - {\bm y})) = i
 \delta^{(3)}(\tilde{\bm x}- \tilde{\bm y})\,, 
\end{align}
as well as
\begin{align}
 &  \left[ \tilde{\zeta}(t,\, \tilde{\bm x}),\, \tilde{\zeta}(t,\, \tilde{\bm y})
 \right] =  \left[ \tilde{\pi}(t,\, \tilde{\bm x}),\, \tilde{\pi}(t,\, \tilde{\bm y})
 \right] = 0\,.
\end{align}

The Hamiltonian density for
$\tilde{\zeta}(\tilde x)$ and $\tilde{\pi}(\tilde x)$ is obtained 
in the standard way as
\begin{align}
\tilde{{\cal H}} \left[
 \tilde{\zeta}(t,\tilde{\bm{x}}),\, \tilde{\pi}(t,\tilde{\bm{x}})
 \right] 
&:= 
\tilde{\pi}(t, \tilde{\bm{x}}) \partial_t
 \tilde{\zeta}(t,\, \tilde{\bm{x}}) 
- 
{\cal L}[(\partial_t\tilde\zeta(t,\, \tilde{\bm{x}}))_{\sbm{x}}, \tilde\zeta(t,\, \tilde{\bm{x}})-s(t)]
 \cr
&= 
\tilde{\pi}(t, \tilde{\bm{x}})  (\partial_t
 \tilde{\zeta}(t,\tilde{\bm{x}}))_{\sbm{x}} - 
{\cal L}
[(\partial_t\tilde\zeta(t,\, \tilde{\bm{x}}))_{\sbm{x}}, \tilde\zeta(t,\, \tilde{\bm{x}})-s(t)]
- \dot{s}(t) \, \tilde{\pi}(t, \tilde{\bm{x}}) \tilde{\bm{x}} 
 \! \cdot \!\partial_{\sbm{\tilde x}}
 \tilde{\zeta}(t,\, \tilde{\bm{x}}) 
\cr
 & = 
{\cal H}
[\tilde\zeta(t,\, \tilde{\bm{x}})-s(t), \tilde\pi(t,\,
 \tilde{\bm{x}})]
  -
 \dot{s}(t) 
\tilde{\pi}(t, \tilde{\bm{x}}) \tilde{\bm{x}} \cdot \partial_{\sbm{\tilde x}}
 \tilde{\zeta}(t,\, \tilde{\bm{x}})~,  \label{Exp:tH}
\end{align}
where in the equality on the second line we used Eq.~(\ref{Exp:temp}). 
The last equality is exactly the same Legendre transformation 
as in the original system and therefore we can use the 
same functional form of the Hamiltonian density ${\cal H}$. 

Assuming that $s(t)$ is as small as $\tilde{\zeta}(x)$ and $\tilde{\pi}(x)$, we
decompose the Hamiltonian densities ${\cal H}$ and $\tilde{\cal H}$ into
the non-interacting parts, which include only the quadratic terms, and the
interacting parts as 
\begin{align}
 & {\cal H}[ \zeta(x),\,\pi(x)] = {\cal
 H}_0[\zeta(x),\, \pi(x)] +{\cal H}_I [\zeta(x),\,\pi(x)]\,,
\end{align}
and
\begin{align}
 & \tilde{\cal H} \left[ \tilde{\zeta}(x),\, \tilde{\pi}(x) \right] = {\cal
 H}_0 \left[ \tilde{\zeta}(x),\, \tilde{\pi}(x)  \right] +
 \tilde{{\cal H}}_I \left[ \tilde{\zeta}(x),\, \tilde{\pi}(x) \right]\,.  \label{Exp:Hdec}
\end{align}
In the above we used the coordinates $\bm{x}$ instead of
$\tilde{\bm{x}}$ for the $\cvt$ system, but 
it will not cause any confusion 
after the relations between the 
$\cv$ and $\cvt$ systems have been established. 
Here, we replaced
${\cal H}_0 [ \tilde{\zeta}(x) -s(t),\, \tilde{\pi}(x)  ]$
with $ {\cal H}_0 [ \tilde{\zeta}(x),\, \tilde{\pi}(x)  ]$, 
since $\zeta(x)$ always 
appears with the spatial derivative in ${\cal H}_0 [\zeta(x),\pi(x)]$. 
Remarkably, the 
non-interacting part of the Hamiltonian density
does not change at all under the dilatation
transformation. 
Using Eq.~(\ref{Exp:tH}), we find that the interaction
Hamiltonian $\tilde{{\cal H}}_I[\tilde{\zeta},\, \tilde{\pi}]$ 
is given by
\begin{align}
 &   \tilde{{\cal H}}_I \left[\tilde{\zeta}(x),\, \tilde{\pi}(x) \right] 
 := {\cal H}_I \left[ \tilde{\zeta}(x) - s(t),\, \tilde{\pi}(x)  \right]  -
 \dot{s}(t)  \tilde{\pi}(x)\, \bm{x}\! \cdot\! \partial_{\sbm{x}}
 \tilde{\zeta}(x) \,.  \label{Def:tH}
\end{align}
In this way, we can write down $\tilde{{\cal H}}_I$ only in terms of 
$\tilde{\zeta}(x)-s(t)$, $\tilde{\zeta}$ with
differentiation, $\tilde\pi$ and $\dot{s}(t)$. 
In Ref.~\cite{SRV1}, we introduced the two sets of the canonical
conjugate variables which are connected by the dilatation transformation
with a constant parameter $s$. When we take the limit where $s(t)$ is
constant, the Hamiltonian density $\tilde{\cal H}(x)$
takes the same functional form as ${\cal H}(x)$ except for the constant
shift of $\tilde{\zeta}(x)$ by $-s$. It is because, without modifying the
gauge condition,  we can perform the dilatation transformation with the
constant parameter $s$ also in the whole universe. Then the action which
preserves the diffeomorphic invariance becomes invariant 
under the change from $\zeta(x)$ to $\zeta(t, e^{-s}\bm{x})-s$.
Here we have extended the argument in Ref.~\cite{SRV1} to
allow $s$ to depend on time.  As we mentioned in Sec.~\ref{Sec:Intro},
this extension plays the crucial role in our discussion about the
secular growth. In the next section, we will show that all the interaction
vertices in the canonical system $\cvt$ are
composed only of the IR irrelevant operator.

\section{Interaction Hamiltonian with the IR irrelevant operators} \label{Sec:Inf}
In this section, we describe the first two of the three items we
raised in Sec.~\ref{Sec:Intro}. In the preceding section, 
we derived the Hamiltonian for the canonical variables $\tilde{\zeta}(x)$ and
$\tilde{\pi}(x)$. Since $\cv$ and
$\cvt$ are connected by the canonical transformation, if we choose the
same initial state in both of the two canonical systems, the
$n$-point functions for the same operator, for instance $\gR$, calculated in these
canonical systems should agree with each other. However, even if we
adopt operationally the same scheme to select the initial state
in these two systems, it does not guarantee that the selected initial
states are the same. In Sec.~\ref{SSec:Euclidean}, after
we describe the definition of the Euclidean vacuum, we will show that
the condition of the Euclidean vacuum operationally selects the
same quantum state irrespective of the choice of the canonical variables. 
This ensures the equivalence of these two
canonical systems including the choice of the initial quantum
state, which we mentioned in the item 2. In
Sec.~\ref{SSec:np}, we will perform the quantization using the canonical
variables $\cvt$. As we will show in Sec.~\ref{RLR}, by virtue of the
equivalence between the two canonical systems, the interaction vertices 
for $\cvt$ can be expressed in terms of operator products composed only
of the IR irrelevant operators.

\subsection{Euclidean vacuum and its uniqueness}   \label{SSec:Euclidean}
In the case with a massive scalar field in de Sitter spacetime,
the boundary condition specified by rotating the time path in 
the complex plane can be understood as 
requesting the regularity of correlation functions on the Euclidean
sphere which can be obtained by the analytic continuation from the ones 
on de Sitter spacetime. The vacuum state thus defined is called Euclidean vacuum
state. Because of the similarity, here we also refer to the state which is
specified by a similar boundary condition as the Euclidean vacuum. To be
more precise, we define the Euclidean vacuum as follows. In the in-in
formalism, the insertion of interaction vertices is ordered along the
closed time path. By rotating the time path toward the imaginary
plane, the forward time evolution begins at
$\eta(t_i)= -\infty(1-i\epsilon)$ and ends at the
final 
time $t_f$ and the backward time evolution begins at $t_f$ and ends at 
$\eta(t_i)=-\infty(1+i\epsilon)$. Here we set $\epsilon$
to a small positive number. Since rotating the time path can be better
understood by using the conformal time $\eta$, we introduced the
conformal time $\eta$ as 
\begin{equation}
\eta(t):=\int^t \frac{\dd t'}{e^{\rho(t')}}=\int^{\rho(t)}
 \frac{\dd \rho'}{e^{\rho'} \dot{\rho}(\rho')}~.
\end{equation}
We define the Euclidean vacuum, requesting the regularity of the
$n$-point functions with an arbitrary natural number $n$ in the limit of 
$\eta(t_i) \to -\infty(1 \pm i \epsilon)$, {\it i.e.,}
\begin{align}
 &  F_n(x_1,\, \cdots x_n) := \langle T_c\, \zeta(x_1) \cdots \zeta(x_n) \rangle < \infty \qquad {\rm as}\,\,
  \quad \eta(t_a) \to  - \infty (1 \pm i \epsilon)\,,  \label{Def:EV}
\end{align}
where $a=1, \cdots n$ and $T_c$ denotes the time ordering along the closed
time path. We first show that the $n$-point functions of $\zeta$ are
uniquely fixed by requesting the condition~(\ref{Def:EV}). In this
paper, for simplicity, we assume that 
$e^\rho \dot\rho(\rho)$ is rapidly increasing in time so that
\begin{align}
  |\eta(t)|={\cal O} \left( 1/e^{\rho(t)} \dot{\rho}(t) \right)\,.
 \label{eta}
\end{align}


Next, we show that the boundary condition of the Euclidean vacuum
uniquely determines the $n$-point functions $F_n(x_1,\, \cdots x_n)$. 
We schematically describe the Heisenberg equation for $\zeta(x)$ as 
\begin{align}
 & {\cal L} \zeta = {\cal S}_{NL}[\zeta]\,, \label{Eq:Sc}
\end{align} 
where ${\cal L}$ is the second-order differential operator:
\begin{align}
 & {\cal L} := \partial^2_\rho + (3- \varepsilon_1 + \varepsilon_2)
 \partial_\rho -\frac{\partial^2}{e^{2\rho} \dot{\rho}^2}~.  \label{Def:cL}
\end{align}
For notational convenience, we introduced the horizon flow functions, 
\begin{align}
 & \varepsilon_1:= -\frac{1}{\dot{\rho}} \frac{\dd}{ \dd \rho} \dot{\rho},\qquad \quad
 \varepsilon_{n}:= \frac{1}{\varepsilon_{n-1}} \frac{\dd}{\dd \rho} \varepsilon_{n-1}~,
\end{align}
with $n\geq 2$, but we do not assume that these functions are small to
keep the background evolution unconstrained except for requesting
Eq.~(\ref{eta}), which is valid, for instance, when $\varepsilon_n$ are
constant in time. Using the Heisenberg equation (\ref{Eq:Sc}), we can obtain the evolution equation
of the path-ordered $n$-point functions $F_n(x_1,\, \cdots x_n)$ as
\begin{align}
 & {\cal L}_{x_a} F_n(x_1,\,\cdots ,\, x_n) = {\cal V}^{(a)}_{NL}[ \{
 F_m\}_{m>n}]\,, 
\label{Eq:zetan0} 
\end{align}
where ${\cal L}_{x_a}$ is the derivative operator
${\cal L}$ given in Eq.~(\ref{Def:cL}) with the coordinates $x$ 
replaced with $x_a$. Since the equation of motion for $\zeta(x)$ is
non-linear, the equation (\ref{Eq:zetan0}) includes the source term
(the right hand side) composed of
$m$-point functions of $\zeta(x)$ with $m>n$. We can verify the
uniqueness of the $n$-point functions for $\zeta(x)$ by showing that
solution of Eq.~(\ref{Eq:zetan0}) is uniquely fixed by the boundary
condition (\ref{Def:EV}). To show this uniqueness, we formally solve the equation (\ref{Eq:zetan0}) as
\begin{align}
 &  F_n(x_1,\,\cdots ,\, x_n) = f_n(x_1,\,\cdots ,\, x_n)
 + {\cal L}^{-1}_{x_a} {\cal V}^{(a)}_{NL}[ \{ F_m\}_{m>n}]\,, 
\end{align}
where $f_n(x_1,\,\cdots ,\, x_n)$ is a homogeneous solution, 
while we assume that the specific solution
${\cal L}^{-1}_{x_a} {\cal V}^{(a)}_{NL}[ \{ F_m\}_{m>n}]$ 
satisfies the regularity condition in the
limits $\eta(t_a) \to  - \infty (1 \pm i \epsilon)$.  
Now the question is whether the boundary condition (\ref{Def:EV}) allows us to add 
any homogeneous solutions. In the Fourier space, 
$f_n$ can be expanded by $e^{-ik\eta(t_a)}$ or
$e^{ik\eta(t_a)}$ in the limits $\eta(t_a) \to - \infty(1 \pm i \epsilon)$.  The
regularity at $\eta(t_a) \to - \infty (1 + i \epsilon)$
accepts $e^{-ik\eta(t_a)}$ only, while the regularity at
$\eta(t_a) \to - \infty (1 - i \epsilon)$
accepts the other. Thus the regularity
condition in the two limits does not allow to add any 
homogeneous solutions $f_n$, which implies that the $n$-point functions
$F_n(x_1,\,\cdots ,\, x_n)$ are uniquely fixed by the boundary
condition of the Euclidean vacuum.

Next, we show that this uniqueness is ensured independent of whether we
use the canonical variables $\cv$ or $\cvt$. 
We employ the boundary condition of the
Euclidean vacuum 
for the canonical variable $\tilde{\zeta}$ as well,
requesting 
\begin{align}
 &  \langle T_c\, \tilde\zeta(x_1) \cdots \tilde\zeta(x_n) \rangle_{\cvt} < \infty \qquad {\rm as}\,\,
  \quad \eta(t_a) \to  - \infty (1 \pm i \epsilon)\,.  \label{Def:EVt}
\end{align}
Then, we can show that the path-ordered $n$-point functions 
\begin{align}
 & \tilde{F}_n(x_1,\, \cdots x_n) := \langle T_c \tilde{\zeta}(t_1,\,
 e^{s(t_1)}\bm{x}_1)  \cdots \tilde\zeta(t_n,\, e^{s(t_n)} \bm{x}_n) \rangle_{\cvt}\,, 
\end{align}
agree with the $n$-point functions 
$
F_n(x_1,\, \cdots x_n)=
\langle T_c\, \zeta(x_1) \cdots \zeta(x_n) \rangle_{\cv}
$ fixed by the
boundary condition (\ref{Def:EV}), {\it i.e.,}
\begin{equation}
\tilde{F}_n(x_1,\, \cdots x_n) = F_n(x_1,\, \cdots x_n).
\label{eq3105}
\end{equation}
Here putting the suffixes $\cv$ or $\cvt$, we denote the canonical
variables used in imposing the boundary condition explicitly. 
We again schematically describe the
Heisenberg equation for $\tilde{\zeta}$ as
\begin{align}
 & {\cal L} \tilde{\zeta}  = \tilde{{\cal S}}_{NL}[\tilde{\zeta}]\,. \label{Eq:Sct}
\end{align}
Since $\zeta(x)$ and $\tilde{\zeta}(x)$ are connected by the canonical transformation, the equation of motion obtained by
operating ${\cal L}$ on
\begin{align}
 & \zeta(x) = \tilde{\zeta}(t,\, e^{s(t)} \bm{x}) = \tilde\zeta(x) + s(t) \bm{x} \cdot
 \partial_{\sbm{x}} \tilde{\zeta}(x) + \cdots \,,
\end{align}
can be recast into Eq.~(\ref{Eq:Sc}) by using Eq.~(\ref{Eq:Sct}). 
A similar argument follows for the equations of
motion for the correlation functions $F_n$ and $\tilde{F}_n$. Using the equation of motion
for the $n$-point functions of $\tilde{\zeta}(x)$, which can be derived
from Eq.~(\ref{Eq:Sct}), we can confirm that an operation of 
${\cal L}_{x_a}$ on 
\begin{align}
 \tilde{F}_n(x_1,\, \cdots x_n) 
 & =   \langle T_c \tilde{\zeta}(x_1)  \cdots \tilde\zeta(x_n) \rangle_{\cvt}
 + s(t_1) \langle T_c \bm{x}_1\cdot \partial_{\sbm{x_1}} \tilde{\zeta}(t_1,\,
 \bm{x}_1)  \cdots \tilde\zeta(t_n,\, \bm{x}_n) \rangle_{\cvt} + \cdots 
\end{align}
leads to 
\begin{align}
 & {\cal L}_{x_a} \tilde{F}_n(x_1,\,\cdots ,\, x_n) =  {\cal V}^{(a)}_{NL}[ \{
 \tilde{F}_m\}_{m>n}]\,. 
\label{Eq:zetantilde} 
\end{align}
This equation takes the same form as the equation of motion
(\ref{Eq:zetan0}). We also note that the boundary condition of the
Euclidean vacuum (\ref{Def:EVt}) implies 
\begin{align}
 & \tilde{F}_n(x_1,\,\cdots,\, x_a\,, \cdots ,\, x_n) < \infty \qquad {\rm as}\,\,
\quad -\eta(t_a) \to  \infty (1 \pm i \epsilon)\,. \label{Exp:BCtz}
\end{align}
The equivalence (\ref{eq3105}) is now transparent, because 
the equations of motion ~(\ref{Eq:zetan0}) and (\ref{Eq:zetantilde}), 
and the boundary conditions~(\ref{Def:EV}) and (\ref{Exp:BCtz}) 
are the same, and the latter specify the solutions of
the former uniquely. 
This equivalence is a distinctive property of 
the Euclidean vacuum\footnote{The uniqueness of 
the Euclidean vacuum becomes intuitively clear 
when the Hamiltonian is time independent and the lowest energy
eigenstate is non-degenerate, because the 
$i\epsilon$ prescription selects the unique ground state of the system.}. 
Here we took the boundary conditions for $n$-point functions 
as the definition of the Euclidean vacuum state, assuming the existence of such a quantum state. 
In Sec.~\ref{sec:iepsilon}, we explain such a Euclidean vacuum, 
if exists, should be the one given by the ordinary $i\epsilon$ 
prescription.

\subsection{Rewriting the $n$-point functions}  \label{SSec:np} 
In this subsection, we rearrange the expression for 
the $n$-point functions of the genuinely \gauge invariant 
variable $\gR$ into a more suitable form to examine the regularity of the IR contributions. 
First, solving the three dimensional geodesic equations, we obtain the relation
between the global coordinates $x^i_{gl}$ and the geodesic normal coordinates
$x^i$ as
\begin{align}
 & x^i_{gl} = e^{-\zeta(t, e^{-\zeta} \sbm{x})}  x^i + \cdots\,,
\label{gnc0}
\end{align}
where the ellipsis means 
the terms which vanish 
when $\zeta(x)$ is spatially homogeneous, {\it i.e.}, the terms suppressed in the IR limit. 
Note that changing the spatial coordinates into the
geodesic normal coordinates also modifies the UV
contributions. Tsamis and Woodard~\cite{Tsamis:1989yu} showed that using the geodesic
normal coordinates can introduce an additional origin of UV
divergence, which may not be able to be
renormalized by local counter terms~\cite{Miao:2012xc}.
It should be clarified whether this issue is a serious problem or not,
but we defer it to a future study. Instead, to keep the UV contributions under control, we replace
$\zeta(x)$ in Eq.~(\ref{gnc0}) with the smeared curvature perturbation
$\Gbz(t)$, {\it i.e.}, 
\begin{align}
 & x^i_{gl} = e^{-\Gbz(t)}  x^i\,, 
\label{gnc}
\end{align}
with 
\begin{align}
 & \Gbz(t) 
:= \frac{\int \dd^3 \bm{x}\, W_{L_t}(\bm{x}) \zeta(t,
 e^{- \Gbz}\bm{x})
 }{\int \dd^3 \bm{x}\, W_{L_t}(\bm{x})}\,, 
\label{Def:Gbz} 
\end{align} 
where $W_{L_t}(\bm{x})$ is a window function which is 
non-vanishing only in the local region $\Sigma_t \cap {\cal O}$. 
We approximate the averaging scale at each time $t$ by the Hubble scale,
{\it i.e.,} 
$L_t  \simeq 1/\{e^{\rho(t)} \dot{\rho}(t)\}$. 
Although $\Gbz$ appears on
the right-hand side of Eq.~(\ref{Def:Gbz}), $\Gbz$ is defined 
iteratively at each order of the perturbation. 
We calculate the $n$-point functions of 
${\cal R}_x \gz(t, \bm{x})$, instead of $\gR$, with 
\begin{align}
 & \gz (t,\, \bm{x}) := \zeta (t,\,   e^{-\Gbz(t)} \bm{x})\,.
\label{Exp:gz}
\end{align}
Here, ${\cal R}_x$ denotes the IR suppressing operator such as 
\begin{align}
\partial_\rho,\quad \frac{\partial_{\sbm{x}}}{e^{\rho(t)}
 \dot{\rho}(t)},\quad \biggl(1 - \frac{\int \dd^3 \bm{x}
 W_{L_t}(\bm{x})}{\int \dd^3 \bm{y} W_{L_t}(\bm{y})} \biggr), \quad 
\cdots\,,
\end{align}
where $x$ is the spacetime coordinates of the field 
on which these operators act. Although ${\cal R}_x\gz(t,\, \bm{x})$ is not 
genuinely \gauge invariant, 
it is still invariant under the dilatation transformation,
which is associated with the dominant IR contributions. In fact, 
since the smeared curvature perturbation $\Gbz(t)$ transforms into
$\Gbz(t) -f$ 
under the dilatation transformation: $\bm{x} \to e^{-f} \bm{x}$ with a
constant $f$, ${\cal R}_x \gz(x)$ is kept invariant under this 
transformation. By contrast, the constant part of $\gz(x)$ can be modified
under the dilatation transformation as
$\gz(x) \to \gz(x)-f$. Since the genuine \gauge invariant variable $\gR(x)$ 
should not be affected by the dilatation transformation, which is a part of
the residual \gauge transformations, $\gz(x)$ 
appears only in the form of ${\cal R}_x \gz(x)$ 
when we express $\gR(x)$ in terms of $\gz(x)$.
As we can compute 
$\gR(x)$ from ${\cal R}_x\gz(x)$, our goal is to prove that the expectation 
values of products of ${\cal R}_x \gz(x)$ 
are IR regular.

First, we calculate the $n$-point functions of $\gz$ without the IR
suppressing operator ${\cal R}_x$:
\begin{align}
 & \langle\, 0| \gz(t_f,\,\bm{x}_1) \cdots \gz(t_f,\,\bm{x}_n) |0\, \rangle. \label{npz}
\end{align}
Using the eigenstates of $\Gbz(t_f)$, $|\,\sH \rangle\Heisenberg$ which satisfy
$\Gbz(t_f) |\,\sH \rangle\Heisenberg =  \sH |\,\sH \rangle\Heisenberg$, we can
construct a unit operator
\begin{align}
 \bm{1} = \int \dd \sH\, |\, \sH \rangle\,\HeisenbergH\langle \sH \,|\,.  
\end{align}
Inserting it into the expression for the $n$-point functions, 
we obtain 
\begin{align}
 \langle\,0| \gz(t_f,\,\bm{x}_1) \cdots \gz(t_f,\,\bm{x}_n) |0\,
 \rangle &= \int \dd \sH \Bigl\langle\,0 \Big| \zeta(t_f,\, e^{-\sH}\bm{x}_1 ) \cdots
 \zeta(t_f,\, e^{-\sH}\bm{x}_n )  \Big|\, \sH \Bigr\rangle\HeisenbergH\Bigl\langle \sH \Big|
 0\, \Bigr\rangle \cr
 & =  \int \dd \sH \Bigl\langle\,0 \Big| \tilde\zeta(t_f,\,\bm{x}_1 ) \cdots
 \tilde\zeta(t_f,\, \bm{x}_n )  \Big|\, \sH \Bigr\rangle\HeisenbergH\Bigl\langle \sH \Big|
 0\,\Bigr\rangle\,.
\end{align}
In the first line 
we could simply replace ${^g\!\bar{\zeta}}(t_f)$ with $\sH$, because
${^g\!\bar{\zeta}}(t_f)$ and $\zeta(t_f,\, \bm{x})$ commute with each
other. 
Since the Heisenberg
picture field $\tilde{\zeta}(t,\, \bm{x})$ is related to the
interaction picture field $\tilde{\zeta}_I(t, \bm{x})$ as
\begin{align}
 \tilde{\zeta}(t,\, \bm{x}) = \tilde{U}^\dagger_I(t) \tilde{\zeta}_I (t,\, \bm{x}) \tilde{U}_I(t)\,, 
\end{align}
where the unitary operator $\tilde{U}_I(t)$ is given by
\begin{align}
 \tilde{U}_I(t) &:= \lim_{\eta(t_i) \to - \infty (1 - i\epsilon) } T \exp \left[-
 i\int^t_{t_i} \dd t \int \dd^3 \bm{x} \, \tilde{\cal
 H}_I\left[\tilde{\zeta}_I(x),\, \tilde{\pi}_I(x) \right] \right]\, \cr
 & =  \lim_{\eta(t_i) \to - \infty (1 - i\epsilon) }
\sum_{n=0}^\infty (-i)^n \int^t_{t_i} \dd t_n
  \int^{t_n}_{t_i} \dd t_{n-1}
\cdots \int^{t_2}_{t_i} \dd t_1  \cr
 & \qquad  \quad \times \int \dd^3 \bm{x}_n 
 \cdots \int \dd^3 \bm{x}_1\, \tilde{\cal H}_I \left[\tilde{\zeta}_I(x_n),\, \tilde{\pi}_I(x_n) \right]
  \cdots 
\tilde{\cal H}_I\left[\tilde{\zeta}_I(x_1),\, \tilde{\pi}_I(x_1) \right]~.
\end{align}
Thus, the $n$-point function can be rewritten as
\begin{align}
 \langle\,0| \gz(t_f,\,\bm{x}_1) \cdots \gz(t_f,\,\bm{x}_n) |0\, \rangle 
 &= \int \dd s \biggl\langle 0 \Bigl\vert
 \bar{T} \exp \left[i \int \dd t\, \dd^3\! {\bm x}\, 
 \tilde{{\cal H}}_I \left[ \tilde{\zeta}(x),\, \tilde{\pi}(x)\right]
 \right] \cr
& 
\times \tilde\zeta_I(t_f,\, \bm{x}_1) \cdots \tilde\zeta_I(t_f,\, \bm{x}_n)
T \exp \left[-i \int \dd t\, \dd^3\! {\bm x}\,  
\tilde{{\cal H}}_I 
\left[ \tilde{\zeta}(x),\, \tilde{\pi}(x) \right] \right] 
\Bigl\vert \sH\Bigl\rangle\HeisenbergH 
\Bigr\langle \sH\Bigr\vert  0
\biggr\rangle~,  \cr
\label{expression1}
\end{align}
where $\bar{T}$ denotes the anti time-ordered product. 
Notice that the interaction Hamiltonian 
$\tilde{{\cal H}}_I$ does not contain 
the second or higher derivative of $s(t)$. 
We construct unit operators, using the eigenstates $|s(t) \rangle$
and $|\dot{s}(t) \rangle$ which satisfy
\begin{align}
 &  \Gbz_I(t) |s(t) \, \rangle = s(t) |s(t) \, \rangle\,, \qquad
  {^g\dot{\bar{\zeta}}}_I(t) |\dot{s}(t) \, \rangle = \dot{s}(t) |\dot{s}(t) \, \rangle
\end{align}
where 
\begin{align}
 & \Gbz_I(t) :=  \frac{\int \dd^3 \bm{x} W_{L_t}(\bm{x}) \zeta_I(t,
 e^{- \Gbz_I(t)}\bm{x})
 }{\int \dd^3 \bm{x} W_{L_t}(\bm{x})}\,  \label{Def:GbzI}
\end{align}
is the smeared interaction picture field. We next replace all $s(t)$ and $\dot{s}(t)$ 
with ${^g\bar{\zeta}}_I(t)$ and 
${^g\dot{\bar{\zeta}}}_I(t)$, respectively, by  
inserting the unit operators;
\begin{equation}
 {\bf 1}=\int \dd s(t)\, \Bigl\vert s(t)\Bigl\rangle \Bigr\langle
  s(t)\Bigr\vert\,, \qquad
 {\bf 1}=\int \dd \dot{s}(t)\, \Bigl\vert \dot{s}(t)\Bigl\rangle \Bigr\langle
  \dot{s}(t)\Bigr\vert\,. 
\end{equation}
To perform this replacement without ambiguity, 
we fix the operator ordering in $\tilde{{\cal H}}_I$ to 
the Weyl ordering, in which 
$\tilde{\zeta}_I(x)- s(t)$ and $\tilde{\pi}_I(x)$ are
symmetrized. Instead of considering the explicit form of the interaction
Hamiltonian, we use a schematic expression of 
$\tilde{{\cal H}}_I$ which is expanded in a power series of
$\dot{s}(t)$ as
\begin{equation}
  \tilde{{\cal H}}_I \left[ \tilde{\zeta}_I(x),\,
		      \tilde{\pi}_I(x)\right] 
 = {\cal H}_I \left[ \tilde{\zeta}_I(x)-s(t),\, \tilde{\pi}_I(x)\right] 
 =\sum_{\alpha=0} \left\{ \dot{s}(t) \right\}^\alpha 
    \tilde{\cal H}_{(\alpha)} \left[\tilde{\zeta}_I(x)-s(t),\,
			 \tilde{\pi}_I(x) \right] \,, 
\end{equation}
although $\alpha$ is at most 1. 
Here, we stress 
that the perturbations $\tilde{\zeta}_I(x)$ and $s(t)$ appear
in the  Hamiltonian density $\tilde{\cal H}_I$ only in the form of
$\tilde{\zeta}_I(x)-s(t)$ or its spatial differentiations. Inserting the unit operators, we obtain
\begin{align}
  &\tilde{{\cal H}}_I \left[  \tilde{\zeta}_I(x),\, \tilde{\pi}_I(x)\right] 
 =\sum_{\alpha=0}  \int \dd s(t) \int \dd \dot{s}(t)
   \left\{ \dot{s}(t) \right\}^\alpha 
    \tilde{{\cal H}}_{(\alpha)} \left[\tilde{\zeta}_I(x)-s(t),\,
 \tilde{\pi}_I(x) \right] 
  \Bigl\vert s(t)\Bigl\rangle \Bigr\langle
  s(t)\Bigr\vert \dot{s}(t)\Bigl\rangle \Bigr\langle
  \dot{s}(t)\Bigr\vert \,.
\end{align}
After we replace $s(t)$ with
${^g\!\bar{\zeta}}_I(t)$, $\Gbz_I(t)$ is located next to the operator 
$|s(t)\rangle\langle s(t)|$. Noticing the fact that $s(t) |\,s(t)\, \rangle$ can be expressed as
\begin{equation}
 s(t) |\,s(t)\, \rangle=  \Gbz_I(t)  |\,s(t)\, \rangle = 
\frac{\int \dd^3 \bm{x}\, W_{L_t}(\bm{x})\,
  \zeta_I(t,\, e^{-s(t)} \bm{x}) }{\int \dd^3 \bm{x}\,
 W_{L_t}(\bm{x})}  |\,s(t)\, \rangle, 
\label{defGbzI}
\end{equation} 
where in the second equality, we replaced $\Gbz_I(t)$ in the argument of
$\zeta_I$ with $s(t)$, 
we use $\Gbz_I(t)$ expressed as 
\begin{align}
 & \Gbz_I(t) = \frac{\int \dd^3 \bm{x}\, W_{L_t}(\bm{x})\,
  \tilde \zeta_I(x) }{\int \dd^3 \bm{x}\,
 W_{L_t}(\bm{x})}~,
\end{align}
instead of the expression given in Eq.~(\ref{Def:GbzI}), 
when we replace $s(t)$ with $\Gbz_I(t)$. 
Using the formula
\begin{align}
 & \left( \tilde\zeta_I(t,\, \bm{x}) - s (t) \right)  {\cal A} |\,s(t)\, \rangle 
 = \left( \tilde\zeta_I(t,\, \bm{x}) - {^g\!\bar\zeta}_I(t) \right)  {\cal A} |\,s(t)\,
 \rangle + \left[ {^g\!\bar\zeta}_I(t),\, {\cal A} \right] |\,s(t)\, \rangle\,,
 \label{formula}
\end{align}
we replace $(\tilde\zeta_I(x) - s(t))$ with 
$( \tilde\zeta_I(x) - {^g\!\bar\zeta}_I(t) )$ one by
one. By induction, the operator ${\cal A}$ is supposed to 
be composed of
$\tilde{\zeta}_I(x)-s(t)$ and $\tilde{\pi}_I(x)$. 
Since $\Gbz_I(t)$ commute with $\tilde{\zeta}_I(x)- s(t)$, 
the non-vanishing commutation relation is only the following:
\begin{align}
 \left[ {^g\!\bar\zeta}_I(t),\,\tilde\pi_I(t, \bm{x}) \right]
 &= \frac{1}{\int \dd^3 \bm{x}~W_{L_t}(\bm{x})} \int \dd^3
 \bm{y}~W_{L_t} (\bm{y}) \left[ \tilde{\zeta}_I(t,\, \bm{x}),\,
 \tilde{\pi}_I(t,\, \bm{y})\right] = i \frac{W_{L_t}(\bm{x})}{\int
 \dd^3 \bm{x}~W_{L_t}(\bm{x})} \,,
\end{align}
where we used 
\begin{align}
 & \left[ \tilde{\zeta}_I(t, \bm{x}),\, \tilde{\pi}_I(t,\, \bm{y}) \right]
 =  \tilde{U}_I(t) \left[ \tilde{\zeta}(t, \bm{x}),\,
 \tilde{\pi}(t,\, \bm{y}) \right] \tilde{U}^\dagger_I(t)
  = i \delta^{(3)}(\bm{x}-\bm{y})\,.
\end{align} 
Since the commutator including $\Gbz_I(t)$ 
yields only a local function, we can conclude that 
operators left after exchanging $s(t)$ with $\Gbz_I(t)$ 
are also composed of $\tilde{\zeta}_I(x)-s(t)$
and $\tilde{\pi}_I(x)$. 
Repeating this procedure, 
we can replace all $s(t)$ with $\Gbz_I(t)$ as
\begin{align}
  &\tilde{{\cal H}}_I \left[  \tilde{\zeta}_I(x),\, \tilde{\pi}_I(x)
 \right] 
=\sum_{\alpha=0}  \int \dd s(t) \int \dd \dot{s}(t)
   \left\{ \dot{s}(t) \right\}^\alpha 
    \tilde{\cal H}_{(\alpha)}' \left[\tilde{\zeta}_I(x)-\Gbz_I(t),\,
 \tilde{\pi}_I(x) \right] 
  \Bigl\vert s(t)\Bigl\rangle \Bigr\langle
  s(t)\Bigr\vert \dot{s}(t)\Bigl\rangle \Bigr\langle
  \dot{s}(t)\Bigr\vert \,,
\end{align}
where to denote the modification after the replacement of $s(t)$ with
$\Gbz_I(t)$, we put $'$ on the interaction Hamiltonian. Replacing 
$\dot{s}(t)$ with ${^g\dot{\bar{\zeta}}}(t)$, we obtain
\begin{align}
  &\tilde{{\cal H}}_I \left[  \tilde{\zeta}_I(x),\, \tilde{\pi}_I(x)
 \right] 
 =\sum_{\alpha=0}  \int \dd s(t) \int \dd \dot{s}(t)\,
    \tilde{\cal H}_{(\alpha)}' \left[\tilde{\zeta}_I(x)-\Gbz_I(t),\,
 \tilde{\pi}_I(x) \right]  
   \Bigl\vert s(t)\Bigl\rangle \Bigr\langle
  s(t)\Bigr\vert \dot{s}(t)\Bigl\rangle \Bigr\langle
  \dot{s}(t)\Bigr\vert \left\{ {^g\dot{\bar{\zeta}}}_I(t) \right\}^\alpha  \,.
\end{align}
We repeat this procedure for all integrating Hamiltonian densities which
appear in the perturbative expansion of the $n$-point functions 
(\ref{expression1}). After these replacements, the possible dependence of the $n$-point functions on $s(t)$ and
$\dot{s}(t)$ remains only in 
$\vert s(t)\rangle\langle s(t)\vert$ and
$\vert \dot{s}(t)\rangle\langle \dot{s}(t)\vert$. Since requesting 
the Euclidean vacuum uniquely determines the initial state independent
of $s(t)$ and $\dot{s}(t)$, we can remove the identity operators
$\int \dd s(t)\, \vert s(t)\rangle\langle s(t)\vert$ and 
$\int \dd \dot{s}(t)\, \vert \dot{s}(t)\rangle\langle \dot{s}(t)\vert$
as long as we choose the Euclidean vacuum.
(From the same argument, we can remove the identity operator
$\int \dd \sH\, \vert \sH \rangle\,\HeisenbergH\langle \sH \vert$.) Then, the Hamiltonian
density is recast into
\begin{align}
  \tilde{{\cal H}}_I \left[ \tilde{\zeta}_I(x),\,
 \tilde{\pi}_I(x)\right]  \to \sum_{\alpha=0}     \tilde{\cal H}_{(\alpha)}' \left[\tilde{\zeta}_I(x)-\Gbz_I(t),\,
 \tilde{\pi}_I(x) \right]  \left\{ {^g\dot{\bar{\zeta}}}_I(t)
 \right\}^\alpha  \,. \label{Exp:NH}
\end{align}
Note that we can express 
${^g\!\dot{\bar{\zeta}}}_I(t)$ as
\begin{align}
  {^g\!\dot{\bar{\zeta}}}_I(t) &=  \int \dd^3 \bm{x}\, \partial_t \left\{
 W_{L_t}(\bm{x}) \over \int \dd^3 \bm{x} W_{L_t}(\bm{x})  \right\}
 \tilde{\zeta}_I(x) +  {\int \dd^3 \bm{x} W_{L_t}(\bm{x})\, \partial_t\tilde{\zeta}_I(x) \over
 \int \dd^3 \bm{x} W_{L_t}(\bm{x})} \cr 
 &=  \int \dd^3 \bm{x}\,  \partial_t \left\{
 W_{L_t}(\bm{x}) \over \int \dd^3 \bm{x} W_{L_t}(\bm{x})  \right\} \left\{  \tilde{\zeta}_{I}(x)
 -\Gbz_I(t) \right\}+ {\int \dd^3 \bm{x} W_{L_t}(\bm{x}) \partial_t\tilde{\zeta}_I(x) \over
 \int \dd^3 \bm{x} W_{L_t}(\bm{x})}\,,
\end{align}
where in the last equality, we inserted
$  0 =
 \Gbz_I(t) \,  \partial_t \left\{ \int \dd^3 \bm{x}
 W_{L_t}(\bm{x}) / \int \dd^3 \bm{x} W_{L_t}(\bm{x})  \right\}\,$
and the last term in the last line can be written in terms of $\tilde\pi_I(x)$.

In this way, we can show that all $\tilde{\zeta}_I$s in the
interaction vertices are multiplied by an IR suppressing operator ${\cal R}_x$. Notice that, replacing the c-number
parameter $s(t)$ with the operator $\Gbz_I(t)$, we rewrote the
Hamiltonian density as in Eq.~(\ref{Exp:NH}). In this procedure, we used
the fact that the initial state specified by the boundary
condition of the Euclidean vacuum does not depend on the choice of the canonical variables. 
We should emphasize that if this equivalence of the initial state 
were not guaranteed, we could not express the interaction
Hamiltonian only in terms of $\tilde{\zeta}_I$s with an IR
suppressing operator.

\subsection{Restricting the interaction vertices to the local region}
\label{RLR}
In the above discussion, we found that the interaction picture fields 
which appear in the interaction vertices 
can be expressed only in terms of $\tilde\pi_I(x)$ 
and $\tilde\zeta_I(x) -{^g\!\bar\zeta}_I(t)$. Now, we can verify 
the item 1 presented in Sec.~\ref{Sec:Intro}, which claims
that the interaction vertices are constructed only from the IR
irrelevant operators. As we showed in the previous subsection,
all the interaction picture fields are associated with 
an IR suppressing operator ${\cal R}_x$, which increases the power law index with respect 
to the wavenumber $k$ in the IR limit.
To complete the proof of the argument given in the item 1, 
we need to show that the inverse Laplacian
$\partial^{-2}$, which appears in solving the constraint equations 
to obtain the lapse function and the shift vector, 
does not reduce the power law index with respect to $k$ 
in the IR limit. The potential danger can be understood as follows. 
When we choose the boundary
condition specified by the regularity at the spatial infinity 
following the standard procedure, the action of the operator $\partial^{-2}$ yields a multiplicative 
factor $1/k^2$. This IR singular behavior arises because the 
information from the outside of our observable region is used to determine 
the lapse function and the shift vector. 

To remove this potential IR singular behavior originating from 
the inverse Laplacian, we need to discuss the causality. The causality is basically 
maintained even at the quantum level in the sense that the 
interaction vertices located outside our observable region 
${\cal O}$ are decoupled in the in-in formalism. In the ordinary 
field theory with a local interaction, this can be shown by systematically replacing 
the Wightman function $G^{+}$ with the retarded Green function 
plus $G^{-}$ (see Appendix of Ref.~\cite{SRV1}). 
However, when the gravitational perturbation is taken into account, 
it becomes less transparent whether the causality is maintained owing to the
issue of the lapse function and the shift vector mentioned above. 

Here, we should recall that what we really need to evaluate is 
the expectation values of genuinely \gauge invariant variables, 
which do not depend on the choice of the residual \gauge 
degrees of freedom.  
As we explicitly showed in Appendix \ref{Sec:constraint}, using the
residual \gauge degrees of freedom, we can modify the boundary
conditions of the lapse function $N$ and the shift vector $N_i$ so that
the terms associated with $\partial^{-2}$ are completely specified 
by the fields within the local region ${\cal O}$. 
Then, the operation of the
non-local operator $\partial^{-2}$ no longer reduces the power law 
index with respect to ${k}$. 
In this way, using the degrees of freedom in the choice of 
boundary conditions, we can localize all the interaction 
vertices within the causally connected local region ${\cal O}$. 
Since ${\cal R}_x \zeta(x)$ is not invariant under the 
residual \gauge transformations, 
their $n$-point functions are not invariant in general 
under the change of boundary conditions of $N$ and $N_i$.
However, when we calculate $n$-point functions for the genuinely
\gauge invariant operator $\gR$ using those for ${\cal
R}_x \gz$, changing the boundary
conditions should not affect the result.

\section{The IR regularity and the absence of the secular growth}   
\label{Sec:IRregularity}
In this section, we will calculate the $n$-point functions of
${\cal R}_{\sbm{x}} \gz(t_f,\, \bm{x})$, 
properly taking into account not only the IR-modes but also 
the modes with $k|\eta(t)| \agt 1$. As stressed in Sec.~\ref{Sec:Intro},
to prove the absence of the secular growth, we need to evaluate the
contribution of the latter modes carefully. In the preceding
section, we showed that, using the canonical variables $\tilde{\zeta}(x)$
and $\tilde{\pi}(x)$, we can expand the $n$-point functions of 
${\cal R}_{\sbm{x}} \gz(x)$
for the Euclidean vacuum only in terms of the IR irrelevant operators. 
In this section, based on the perturbative expansion in the $\cvt$ system, we 
will discuss the IR regularity and 
the absence of secular growth in the $n$-point functions. 

For our current discussion,  
the explicit form of the interaction Hamiltonian density 
$\tilde{\cal H}_I$ is not necessary.  
We use a formal expression 
\begin{align}
 & \tilde{\cal H}_I[\tilde{\zeta}_I(x),\, \tilde{\pi}_I(x)] = \Mp^2 e^{3\rho} \dot{\rho}^2 \varepsilon_1(t) \sum_{n=3} \lambda(t)
 \prod_{m=1}^n {\cal R}^{(m)}_x \tilde{\zeta}_I(x)
 \,,
\end{align}
where $\lambda(t)$ is an ${\cal O}(1)$ dimensionless time dependent function which can be expressed
only in terms of the horizon flow functions. To discriminate different IR suppressing operators, 
we associate a superscript $(m)$ on ${\cal R}_x$.

\subsection{Euclidean vacuum as is obtained by $i\epsilon$ prescription}
\label{sec:iepsilon}
In the preceding section, we introduced the Euclidean vacuum 
as a vacuum state which satisfies the boundary condition (\ref{Def:EV})/(\ref{Def:EVt}). 
Here we show that this condition forces us to adopt the ordinary perturbative description of
the $i\epsilon$ prescription. We expand the curvature perturbation $\tilde{\zeta}_I(x)$ as
\begin{align}
 \tilde\zeta_I(x) &= \int \frac{\dd^3 \bm{k}}{(2\pi)^{3/2}}
    \,e^{i\sbm{k}\cdot\sbm{x}}  v_k(t) 
\tilde{a}_{\sbm{k}}+ {\rm h.c.}\,, \label{Exp:zeta} 
\end{align}
where $\tilde a_{\sbm{k}}$ is the annihilation operator, which satisfies
\begin{align}
 & \left[\, \tilde{a}_{\sbm{k}},\, \tilde{a}_{\sbm{k}'}^\dagger \,\right] = \delta^{(3)}
 (\bm{k} - \bm{k}')\,, \qquad \left[\, \tilde{a}_{\sbm{k}},\, \tilde{a}_{\sbm{k}}\,
 \right]  =0\,.
\end{align}
The mode function $v_k(t)$ should satisfy
\begin{align}
 & \left[ \partial_\rho^2 + (3- \varepsilon_1 + \varepsilon_2)
 \partial_\rho + \left(
{k\over e^\rho \dot\rho}\right)^2 \right] v_k = 0~.
\end{align}
Since the boundary condition (\ref{Def:EV})/(\ref{Def:EVt}) should hold at the tree
level, the asymptotic form of the positive frequency mode function $v_k(t)$ should be 
$\propto e^{-ik\eta(t)}$. Factoring out this time dependence at $\eta\to -\infty$, 
we express $v_k(t)$ as 
\begin{align}
 & v_k(t) = 
\frac{{\cal A}(t)}{k^{3/2}} f_k(t) e^{-i k\eta(t)}\,, \label{Exp:vk}
\end{align}
where we introduced 
\begin{align}
 & {\cal A}(t) := \frac{\dot{\rho}(t)}{\sqrt{\varepsilon_1(t)} \Mp}\,,
\end{align}
as an approximate amplitude of the fluctuation.
The function $f_k(t)$ satisfies the regular second order differential 
equation with the boundary condition 
\begin{equation}
 f_k(t)\to {k\over \sqrt{2}\,e^\rho \dot\rho}\qquad 
\mbox{for}\quad -k\eta(t) \to \infty~.  
\end{equation}
Since both the differential equation and the boundary condition of $f_k(t)$
are analytic in $k$ for any $t$, the resulting function should be
analytic as well. Namely, $f_k(t)$ does not have any singularity
such as a pole on the complex $k$-plane. We suppose that a positive frequency
function for a general vacuum except for the Euclidean vacuum is given by a linear
combination of $v_k$ and $v_k^*$ with the Bogoliubov coefficients which have some nontrivial 
structure of singularities in the complex $k$-plane 
or diverge at infinity. The only exception to evade the singularity is
setting the Bogoliubov coefficients to constants, but then the UV behavior does not agree with 
the one in the Minkowski vacuum.

On the other hand, in the limit $-k\eta(t_k) \ll 1$, the
function $f_k(t)$ is proportional to $
 {{\cal A}(t_k)/ {\cal A}(t)},
$
where $t_k$ is the Hubble crossing time defined by 
$-k\eta(t_k)=1$, because the curvature perturbation should be constant in this limit. 
Hence, the expansion for small $k$ is in general given by  
\begin{equation}
 {\cal A}(t)f_k(t)= {\cal A}(t_k)\left[1+{\cal O} (k|\eta(t)|)\right]~.
\label{asymptoticexp}
\end{equation}
By using Eq.~(\ref{Exp:vk}), the Wightman function is given by
\begin{align}
  G^+(x,\, x') &= \int \frac{\dd^3 \bm{k}}{(2\pi)^3} e^{i \sbm{k} \cdot
 (\sbm{x} - \sbm{x}')} v_k(t) v^*_k(t')  \cr
 &= {\cal A}(t) {\cal A}(t') \int \frac{\dd^3 \bm{k}}{(2\pi)^3} \frac{1}{k^3} e^{i \sbm{k} \cdot
 (\sbm{x} - \sbm{x}')} f_k(t) f^*_k(t') e^{ik
 (\eta(t')-\eta(t))} \,.
\end{align}

\begin{figure}[t]
\begin{center}
\begin{tabular}{cc}
\includegraphics[width=14cm]{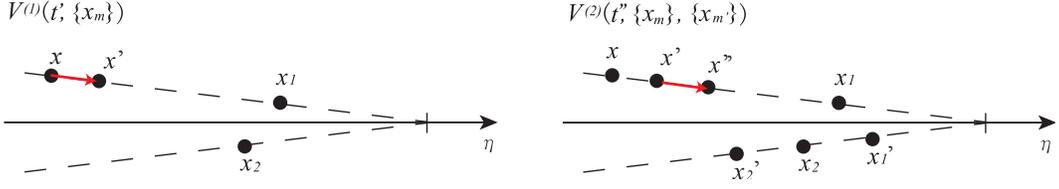}
\end{tabular}
\caption{We perform the vertex integrations from the closer vertices to
 the future or past end of the closed time pass. The left figure
 represents the integration about the vertex which is closest to the
 past end and the right figure represents the integration about the next
 to the closest to the past end.}  
\label{Fg:V}
\end{center}
\end{figure}

Using the in-in formalism, the $n$-point functions can be expanded by
the Wightman function. At this point, the vertex integrals should
start with $\eta=-\infty$ to be able to impose the boundary condition of the
Euclidean vacuum (\ref{Def:EV})/(\ref{Def:EVt}). 
Although the integrands of the vertex integrals are infinitely oscillating 
in the limit $\eta\to -\infty$, the time integration can be made
convergent by adding a small imaginary part to the time coordinate, which is nothing but the 
ordinary $i\epsilon$ prescription.
To see the convergence of the time integration more explicitly, we first consider the integral for the 
vertex which is closest to the past infinity 
$\eta \to -\infty  (1-i\epsilon)$ (see Fig.~\ref{Fg:V}). 
The interaction picture fields $\tilde{\zeta}_I(x)$ included in this
vertex are contracted with $\tilde{\zeta}_I(x_m)$ 
contained in vertices labelled by $m=1, 2, \cdots, n$, and give the
Wightman function $G^+(x_m,\, x)$. Then, the vertex
integration with $n$ interaction picture fields is given by
\begin{align}
 V^{(1)}(t',\, \{x_m\})&:= \Mp^2 \int^{t'}_{t_i} \dd t \int \dd^3 \bm{x}\,
 e^{3\rho(t)}\varepsilon_1(t) \dot{\rho}(t)^2  \lambda(t)
 \prod_{m=1}^n {\cal R}_{x_m} {\cal R}_{x}^{(m)} G^+(x_m, x)~.
\label{V1vertex}
\end{align}
The Euclidean vacuum condition (\ref{Def:EV})/(\ref{Def:EVt}) 
requires the convergence of this integral 
when we send $\eta(t_i)\to -\infty$. 
Since the Wightman functions contain 
the exponential factor $e^{i \eta(t){\sum}_m k_m}$, 
the integral can be made convergent by changing the integration contour 
as shown in the left panel of Fig.~\ref{Fg:V}, 
which is exactly what is known as the $i\epsilon$ prescription.

The vertex integration next to the closest to
the past infinity 
\begin{align}
 &V^{(2)}(t'', \{x_m\}, \{x_{m'}\}) \cr
 & \quad := \Mp^2 \int^{t''}_{t_i} \dd t' \int \dd^3 \bm{x}' e^{3\rho(t')}
 \varepsilon_1(t') \dot{\rho}(t')^2 \lambda(t')
 \prod_{m'=1}^{n'} {\cal R}_{x_{m'}} {\cal R}_{x'}^{(m')} G^+(x_{m'},
 x')  V^{(1)}(t',\, \{x_m\})
\,,  \label{Exp:V2}
\end{align}
can be done in a similar manner, where 
$n'$ is the number of propagators connecting between this second 
vertex and the vertices 
other than the first one. 
If we assume the integration over the time coordinate of the 
first vertex $t$ up to $t'$,  
the exponential factor in $G^+(x_m, x)$ can be replaced as 
\begin{align}
 & e^{i k_m (\eta(t)-\eta(t_m))}\,\,
 \to 
 e^{i k_m (\eta(t')-\eta(t_m))}\,. 
\end{align}
Therefore all the Wightman functions connecting the vertices 
at $t'$ or before $t'$ with the vertices after $t'$ give an 
exponential factor which is suppressed by adding $+i\epsilon$ to 
$\eta$. This is again consistent with the boundary condition of the
Euclidean vacuum. The same argument can be made for the other vertices
as well.

\subsection{The IR/UV suppressed Wightman function}       
\label{SSec:Wightman}
Since all $\tilde{\zeta}_I(x)$s in the interaction
Hamiltonian are multiplied by the IR suppressing operators 
${\cal R}_x$, the $n$-point function of ${\cal R}_x \gz(x)$ can be
expanded by the Wightman function 
${\cal R}_x {\cal R}_{x'} G^+(x,\,x')$ and its complex conjugate
${\cal R}_x {\cal R}_{x'} G^-(x,\, x')$. 
In this subsection, we calculate the Wightman functions multiplied by
the IR suppressing operator, ${\cal R}_x {\cal R}_{x'} G^+(x,\, x')$ for
$t>t'$. 
After integration over the
angular part of the momentum, the Wightman function 
${\cal R}_x {\cal R}_{x'} G^+(x,\,x')$ can be expressed as
\begin{align}
  {\cal R}_x {\cal R}_{x'} G^+(x,\, x') 
 &= {1 \over 2\pi^2}
  \int^{\infty}_0 \frac{\dd
  k}{k}\,  {\cal R}_x{\cal R}_{x'} {\cal A}(t) f_k(t)
{\cal A}(t') f_k^*(t')
\left[{e^{i k \sigma_+(x, x')} -e^{i k \sigma_-(x, x')} \over
  i k (\sigma_+(x, x') -  \sigma_-(x, x'))}
\right]\,, 
\label{Exp:DG+}
\end{align}
where we introduced
$$
\sigma_{\pm}  (x,\, x') :=  \eta(t')-\eta(t) \pm  |\bm{x} - \bm{x}'|\,. 
$$

We first show the regularity of the $k$ integration in
Eq.~(\ref{Exp:DG+}). Since the function $f_k(t)$ is not singular, the
regularity can be verified if the integration converges both in the IR
and UV limits. The regularity in the IR limit is guaranteed by the
presence of the IR suppressing operator. The IR suppressing 
operators ${\cal R}_x$ add at least one extra factor of 
$k|\eta(t)|$ or eliminate the leading
$t$-independent term in the IR limit, and yield
\begin{align}
 {\cal R}_x  {\cal A}(t) f_k(t) \left[{
e^{i k \sigma_+(x, x')} -e^{i k \sigma_-(x, x')} \over
  i k (\sigma_+(x, x') -  \sigma_-(x, x'))} \right] 
  &={\cal A}(t_k) e^{i k \eta(t')} \, {\cO} \left( k |\eta(t)| \right)
\cr
&= {\cal A}(t) e^{i k \eta(t')} \, {\cal O}\left( \left\{  k |\eta(t)|
 \right\}^{(n_s+1)/2} \right)  .  \label{Eq:order}
\end{align}
where we have introduced the spectral index 
$n_s-1:=\dd \log(|{\cal A}(t_k)|^2)/\dd \log k$. Thus, the operation of
${\cal R}_x$ makes the $k$ integration in Eq.~(\ref{Exp:DG+}) regular in
the IR limit. Next, we consider the convergence in the UV limit. In
Eq.~(\ref{Exp:DG+}), the integration contour of $k$ should be
appropriately modified at $k\to\infty$ so that the integral becomes convergent. 
This modification of the integration contour can be also understood 
as a part of the $i\epsilon$ prescription, because adding 
a small imaginary part to all the time coordinates as
$\eta\to \eta\times (1-i\epsilon)$
leads to the replacement 
$\eta(t')-\eta(t)\to \eta(t')-\eta(t)+ i\epsilon$, 
where we note $\eta(t')-\eta(t) < 0$, and hence to introducing an exponential suppression factor 
for large $k$. 
This UV regulator makes the integral finite for the large $k$
contribution except for the case 
$\sigma_{\pm}(x, x') = 0$, where 
$x$ and $x'$ are mutually light-like. Since the expression of the
Wightman function obtained after the $k$ integration is independent of
the value of $\epsilon$, the regulator makes the UV contributions
convergent even after $\epsilon$ is sent to zero. For $\sigma_{\pm}(x, x') = 0$, 
the integral becomes divergent in the limit $\epsilon\to 0$, 
but the divergence related to the behavior of the Wightman functions 
in this limit is to be interpreted as the ordinary 
UV divergences, whose contribution to the vertex integrals 
must be renormalized by 
introducing local counter terms. Thus, the Wightman function
${\cal R}_x {\cal R}_{x'} G^{\pm}(x,\, x')$ is now shown to be a regular
function.

Since the amplitude of the Wightman function with the IR suppressing
operator is bounded from above, we can show the regularity of the $n$-point
functions, if the non-vanishing support of the integrands of the vertex
integrals is effectively 
restricted to a finite spacetime region. Since the causality has been
established with the aid of the residual gauge degrees of freedom, the
question to address is whether vertexes at the distant past is shut off or not. To address the presence of
such a long-term correlation, we discuss the asymptotic behavior of 
the Wightman function ${\cal R}_x {\cal R}_{x'} G^{\pm}(x,\, x')$,
sending $t'$ to a distant past. Recall that when $\sigma_{\pm}(x,x')\ne 0$, 
we can rotate the integration contour 
with respect to $k$ even toward the direction parallel 
to the imaginary axis. Rotating the 
direction of the path appropriately depending on the sign of
$\sigma_{\pm}(x,x')$, the integrand 
becomes an exponentially decaying function of $k$. This rotation of the 
integration contour can 
be done without hitting any singularity in the complex $k$-plane, because the function $f_k(t)$ is guaranteed to be analytic by
construction. If we choose other vacua, this operation induces 
extra contributions from singularities.  Since we send $t'$ to the past
infinity, assuming $|\eta(t')| \gg |\eta(t)|$, $\sigma_\pm(x,x')$ is
$\cO(|\eta(t')|)$, except for the region where the two points are
mutually light-like~\footnote{Let's introduce a physical length scale 
$\lambda_{\rm UV}$ to remove the contributions from the vicinity of the lightcone. 
On the time slice specified by $\eta'$, 
we neglect the region within the distance 
$\lambda_{\rm UV}$ from the intersection of the 
light cone emanating from $x$ with this time slice. 
Under this restriction, 
we have $\sigma_+(x,x')>|\eta(t')|H(t')\lambda_{\rm UV}$ and 
hence $\sigma_+(x,x')$ turns out to grow in proportion to $|\eta(t')|$. 
This argument might be too heuristic, 
but we believe that the contribution from the region neglected here 
will not change our discussion about the IR regularity of the $n$-point
functions. In order to clarify this point, 
it would be necessary to incorporate the discussion about 
the UV renormalization, which is beyond the scope of this paper.}. Then, the integration of $k$ on the right-hand side of
Eq.~(\ref{Exp:DG+}) is totally dominated
by the wavenumbers with $k \alt 1/|\eta(t')| \ll 1/|\eta(t)|$.
Using Eq.~(\ref{Eq:order}) which gives the asymptotic expansion in the
limit $k |\eta(t)| \ll 1$, we obtain
\begin{align}
  {\cal R}_x {\cal R}_{x'} G^+(x,\, x') 
 &= {\cal A}(t)  {\cal O} \left[ 
  \int^{\infty}_0 \frac{\dd k}{k} \left\{  k \over e^{\rho(t)} \dot{\rho}(t)
 \right\}^{(n_s+1)/2} {\cal R}_{x'} {\cal A}(t') f_k^*(t')
 e^{i k \eta(t')}  \right] \cr
 & = {\cal A}(t) {\cal A}(t') 
 {\cO}\left(\left( |\eta(t)| \over |\eta(t')| \right)^{
 {n_s+1 \over 2}}\right)
\label{OoM} , 
\end{align}
where on the second equality, we performed the $k$ integration, rotating
the integration contour. We should emphasize that we did not employ the
long wavelength approximation regarding the Hubble scale at $t'$ to
properly evaluate the modes $k$ of ${\cal O}(1/|\eta(t')|)$ as well.

\subsection{The secular growth}
\label{SSec:regularity}
In this subsection, focusing on the long-term correlation, we discuss
the convergence of the vertex integrals of the $n$-point functions for
the Euclidean vacuum. 
\begin{figure}[t]
\begin{center}
\begin{tabular}{cc}
\includegraphics[width=13cm]{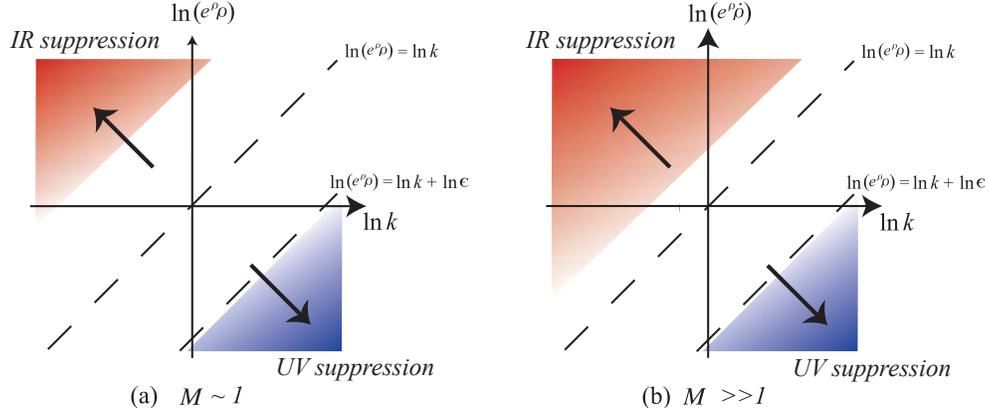} 
\end{tabular}
\caption{These figures show which modes can contribute to the loop
 integrals in the $n$-point function of $\gz$ for the Euclidean
 vacuum. The horizontal axis represents the wavenumber $\ln k$ and the
 vertical axis represents the time 
$\ln (e^{\rho}\dot{\rho}) \simeq \ln(1/|\eta|)$, which
 becomes the number of e-folding in the limit $\varepsilon_1 \ll 1$. The
 red region is suppressed because of the operation of the IR
 suppressing operator ${\cal R}_x$ and the blue region is suppressed
 because of the exponential suppression of the $i\epsilon$
 prescription. The dotted line with $\log (e^\rho \dot{\rho})=\log k$ is the mode of the
 Hubble scale. The left figure (a) is for the case with 
$ M \sim 1$ and the
 right figure (b) is for the case with $M \gg 1$ .}  
\label{fg:S}
\end{center}
\end{figure} 
We start with the integration of the $n$-point interaction 
vertex which is the closest to $\eta=- \infty(1-i\epsilon)$.
By inserting the expression of the Wightman function ${\cal R}_x {\cal
R}_{x'} G^+(x,\, x')$ with $t \gg t'$, given in Eq.~(\ref{OoM}) into 
Eq.~(\ref{V1vertex}), the vertex integral $V^{(1)}$ can be estimated as 
\begin{align}
  & V^{(1)}(t',\,\{x_m\}) ={\cO}\!\left[ \Mp^2
 \int^{t'}_{t_i} \dd t \int \dd^3 \bm{x}\,
 e^{3\rho(t)}\varepsilon_1(t) \dot{\rho}(t)^2  \lambda(t) 
\{{\cal A}(t)\}^n 
\prod_{m=1}^n {\cal A}(t_m) \left( \eta(t_m)  \over \eta(t) \right)^{{n_s+1 \over 2}} \right]\,.
\end{align}
As we have explained in Sec.~\ref{RLR}, 
the interaction vertices are confined within the
observable region, {\it i.e.}, the non-vanishing support of 
the integrand is bounded by $|\bm{x}| \alt L_t \simeq |\eta(t)|$. Thus, we obtain
\begin{align}
& V^{(1)}_n(t',\,\{x_m\}) = {\cO}\left[
 \int^{\eta(t')}_{-\infty}  \frac{\dd \eta}{\eta} \lambda(\eta)  \{{\cal
 A}(\eta)\}^{n-2} \prod_{m=1}^n {\cal A}(t_m)  \left( \eta(t_m)  \over \eta \right)^{{n_s+1 \over 2}} \right] \,.  \label{Exp:V1f}
\end{align}
As we have performed momentum integral first, 
the exponential suppression for large $|\eta|$ is not remaining any more. 
However, picking up $\eta$-dependence of the integrand 
of Eq.~(\ref{Exp:V1f}), we still find that the contribution 
from the distant past is suppressed if 
\begin{align}
 &  \left| \lambda(\eta) \left\{ {\cal
 A}(\eta)\right\}^{n-2}  \eta^{-{n(n_s+1) \over 2} }  \right| \to 0
 \qquad {\rm as} \quad  \eta \to - \infty\,.  \label{Exp:Cond}
\end{align}
Then, the time integral converges, and 
the amplitude of $V_n^{(1)}(\eta',\,\{x_m\})$ is 
estimated by the value of the integrand at 
the upper end of the integration as
\begin{align}
 &  V^{(1)}(t',\,\{x_m\}) =
 {\cO}\left[
 \lambda(t') \{{\cal A}(t')\}^{n-2}
 \prod_{m=1}^n {\cal A}(t_m)  \left( \eta(t_m)  \over \eta(t') \right)^{{n_s+1 \over 2}} \right]
\,.
\end{align}
Therefore, when a Wightman propagator
is connected to a vertex located in the future of $x'$, 
 {\it i.e.}, when $t_m>t'$, the $t$-integration yields the suppression factor 
$\{\eta(t_m)/ \eta(t') \}^{{n_s+1\over2}}$.   
We denote the number of such propagators by $\tilde n$.

Similarly, we can evaluate the amplitude of $V^{(2)}$ as 
\begin{align}
 &V^{(2)}(t'',\, \{x_m\}, \{x_{m'}\}) \cr
 &\quad =\cO\left[
   \int^{\eta(t'')}_{-\infty} \frac{\dd \eta'}{\eta'} \lambda'(\eta')  \{{\cal A}(\eta')\}^{n'-2}
 \prod_{m'=1}^{n'}  {\cal A}(t_{m'})  \left( \eta(t_m) \over \eta' \right)^{{n_s+1 \over 2}} 
\,
 V^{(1)}(t(\eta'),\,\{x_m\})\right] \,. 
\end{align}
Extracting the $\eta'$-dependent part in the above expression, 
we obtain  
\begin{align}
   \int^{\eta(t'')}_{-\infty} \frac{\dd \eta'}{\eta'} \lambda(\eta') \lambda'(\eta')  
  \{{\cal A}(\eta')\}^{n+n'-4} |\eta'|^{-{n_s+1\over 2}(n'+\tilde n)}\,. 
\end{align}
Notice that all the Wightman propagators which are connected to the
field $\tilde{\zeta}_I$ located in the future of $x'$ yield
the suppression factor $|\eta(t'')|^{-{n_s+1\over 2}}$.

Now the generalization becomes easy. For the $N_v$-th vertex, the temporal integration becomes 
\begin{equation}
    \int \frac{\dd \eta_{N_v}}{\eta_{N_v}} \hat\lambda(\eta_{N_v})  \{{\cal
     A}(\eta_{N_v})\}^{N_f-2N_v} |\eta_{N_v}|^{-{n_s+1\over 2}M}\,,
\end{equation}
where $N_f$ denotes the number of $\tilde{\zeta}_I$s contained in the 
vertices up to the ${N_v}$-th, $M$ denotes the number of the Wightman 
propagators connected to a vertex with $\eta>\eta_{N_v}$, and
$\hat{\lambda}$ denotes the product of the interaction
coefficient up to the $N_v$-th vertex. Thus, the convergence condition
is given by
\begin{align}
 &  \left| \hat\lambda(\eta)  \{{\cal
 A}(\eta)\}^{N_f-2N_v} \eta^{-1-{n_s+1\over 2}M}
 \right| \to 0  \qquad {\rm as} \quad  \eta \to - \infty\,. \label{Exp:Ccond}
\end{align}
Since all interaction
vertices have at least one Wightman propagator connected with their
future vertices, $M$ should satisfy $M \geq 1$.

As a simple example, we consider the case where $\varepsilon_1$ is
constant. In this case, $\hat{\lambda}$ is expressed only in terms of
$\varepsilon_1$ and takes a constant value. By assuming $M=1$ and using
$n_s-1 = - 2\varepsilon_1$, the convergence condition yields 
\begin{align}
  -  \varepsilon_1 N + (1- \varepsilon_1)^2 > 0~, 
 \label{Exp:Ccondg2}
\end{align}
with $N:=N_f-2N_v$. In the slow roll limit $\varepsilon_1 \ll 1$,
the above condition is recast into
\begin{equation}
  N  < {\cal O}(1/\varepsilon_1)\,. \label{Exp:Ccondg}
\end{equation}

The intuitive understanding of the above suppression mechanism is 
as follows. In the Euclidean vacuum case, 
only the contributions around the Hubble
scale at each time are left unsuppressed (as shown in Fig.~\ref{fg:S}). 
When only the modes around the Hubble scale, {\it i.e.}, 
$k|\eta| \simeq k/e^{\rho} \dot{\rho}  = {\cal O}(1)$, are relevant, 
the Wightman function ${\cal R}_x {\cal R}_{x'} G^+(x,\, x')$ 
is necessarily suppressed when $\eta(t)/\eta(t') \ll 1$. 
This is because if $x$ and $x'$
are largely separated in time, any Fourier mode 
in the Wightman function cannot be of order of the Hubble scale 
simultaneously at $t$ and $t'$. 
When we consider the contribution of vertices located far in the past,  
at least one Wightman function should 
satisfy $\eta(t)/\eta(t') \ll 1$, and therefore 
it is suppressed.  
However, when we consider a diagram for which a 
cluster of vertices in a distant past is connected to 
the vertices around the observation time by a single 
propagator, {\it i.e.,} in the case with $M=1$, the IR suppression comes only from this 
propagator. 
When the number of operators in the cluster of vertices in 
the past is sufficiently large, the
suppression due to this propagator can be overwhelmed by 
the large amplitude of the fluctuation, which increases as 
the energy scale of inflation increases in the past direction. 
This corresponds to the case when the condition
(\ref{Exp:Ccond}) is broken. However, we should also stress that the
contributions from the distant past are suppressed and the secular
growth never appear in the slow roll inflation, 
unless the order of perturbative expansion 
$N$ takes an extremely large
value such as 
$1/\varepsilon_1 \simeq {\cal O}(10^2)$. 
When the convergence condition (\ref{Exp:Ccond}) is satisfied, 
all the time integrations are dominated by the contributions near its
upper end. The order of magnitude of the $n$-point functions of 
${\cal R}_x \gz(t_f,\,\bm{x})$ is then given by 
\begin{align}
 & \langle\,0| {\cal R}_{x_1} \gz(t_f,\, \bm{x}_1) {\cal R}_{x_2}
  \gz(t_f,\, \bm{x}_2) \cdots  {\cal R}_{x_n}  \gz(t_f,\,
 \bm{x}_n) |0\, \rangle  \simeq \hat{\lambda}(t_f) \{ {\cal A}(t_f)
 \}^{N}~. 
\end{align}

\section{Conclusion and Discussion} \label{Sec:C}
\subsection{Euclidean vacuum satisfies the strong constraint on the initial states}

\begin{figure}[t]
\begin{center}
\begin{tabular}{cc}
\includegraphics[width=12cm]{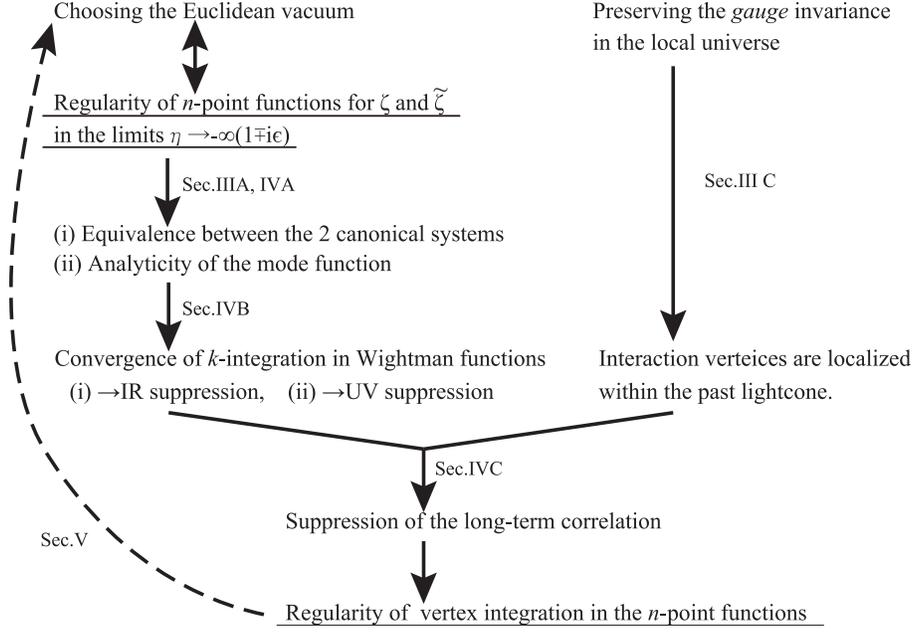}
\end{tabular}
\caption{The outline of the proof which shows the regularity of the
 $n$-point functions of the genuinely \gauge invariant variable for the
 Euclidean vacuum. Since we have left a possibility that the $n$-point
 functions can become regular without requesting the boundary condition
 of the Euclidean vacuum, we used the dotted arrow.}   
\label{Fg:logic}
\end{center}
\end{figure}

In this paper, we showed that when we choose the Euclidean vacuum as the
initial state, the vertex integration in the $n$-point functions for
the genuinely \gauge invariant curvature perturbation 
is regular unless a very high order in the 
perturbative expansion is concerned.
Figure~\ref{Fg:logic} shows the outline of the proof. 
We should emphasize that the regularity of the $n$-point functions in the limits 
$\eta \to - \infty (1 \pm i \epsilon)$ plays a crucial role in the
proof: (i) Requesting this regularity guarantees the
equivalence between two quantum systems, {\it i.e.,} the original system in which the Hamiltonian
contains the IR relevant operators and the quantum system in which the Hamiltonian is totally composed of IR 
irrelevant operators. (ii) It guarantees the analyticity of the
mode function $v_k(t)$ with respect to the wavenumber $k$ for arbitrary 
$t$. By virtue of the aspect (i), we can rewrite the $n$-point functions
of $\gz$ into those expressed in $\cvt$, in which all the field operators 
are manifestly 
associated with the IR suppressing operators, ${\cal R}_x$. 
The aspect (ii) leads to the exponential suppression in
the UV so that the non-vanishing support of the $k$-integration is restricted to 
$- k \eta \alt {\cal O}(1)$. 
It might be intriguing that choosing the Euclidean vacuum
plays the crucial role in discussing the suppressions both in 
the IR and UV components. Since these suppressions make the Wightman
function (in the position space) associated with an IR suppressing operator
regular everywhere except for the light cone limit, 
the missing piece to prove the regularity of the $n$-point
functions is to show that the integration region of each vertex integral
is effectively confined to a finite portion of the spacetime. Using the
residual \gauge degrees of freedom, we can confine the interaction
vertices within the past light cone. Since the long-term correlation is
shut off because of the suppression both in the IR and UV, the
integration region of the vertex integrals is ensured to be
effectively finite. Therefore, the $n$-point functions for the 
Euclidean vacuum are expressed by integrals whose 
integrand and integration region are both finite, and hence 
they are manifestly regular. 
Thus, we conclude that the Euclidean vacuum is a suitable
initial state of the universe which is free from the IR pathology
even in the presence of non-linear interactions.

In this section, we further address the converse question; ``When we
request that the $n$-point functions are finite and free from 
the secular growth, is the Euclidean
vacuum the unique possible initial quantum state?'' 
To be precise, the condition we impose here is 
the regularity of $n$-point
functions on the real time axis including the distant past, {\it i.e.,} $-\eta \gg 1$.  
We naively expect that in this case, the Euclidean vacuum is the unique possibility. 
Since any excitations are blue shifted at an earlier time, 
any small deviation from the Euclidean vacuum state 
at a finite time will lead to 
some singular behavior in the limit $-\eta \gg 1$. 
However, we do not have any rigorous proof about this argument yet. 
There might be a fundamental obstacle when we try to make 
this statement precise. When we trace back the history of the universe, it 
should inevitably enter the regime in which the background energy 
density and hence the amplitude of the vacuum fluctuation are
so high that the perturbative analysis would not make 
sense any more.

As an alternative setup of the problem is to require the 
regularity of the $n$-point functions just for $\eta>\eta_i$ 
with a certain initial time, $\eta_i$.
The relaxed requirement of the regularity allows us to take other
states, if correlation functions for these states can be reinterpreted as correlation
functions for the Euclidean vacuum.
We introduce a new operator 
$$
 A_{(m)}^\dag=\int \dd^3 \bm{x}\, 
 W_{L_t}(\bm{x}){\cal R}^{(m)}_x \tilde \zeta_I(x)~,
$$
with an arbitrary choice of the IR suppression operator ${\cal R}^{(m)}$ 
where $m$ is just the label for distinction. 
Then, we can define 
the 1 particle state by 
$|\,1_{(m)}\rangle
:={\cal N}A_{(m)}^\dag|0\rangle$ with an appropriate 
normalization factor ${\cal N}$. 
The $n$-point functions of ${\cal R}_x\gz(x)$ at the initial time $\eta=\eta_i$ 
for the 1 particle state $|\,1_{(m)}\rangle$ defined at the
initial time can be expressed in terms of the $(n+2)$-point 
functions for the products of ${\cal R}_x\gz(x)$
for the Euclidean
vacuum. When the initial distribution is regular, as we showed in
Ref.~\cite{IRsingle}, the distribution at late times will be kept regular as
well. Similarly, we can construct excited states with plural 
particles. (Similar excited states are discussed in de Sitter
spacetime in Ref.~\cite{MMS}.) 
However, here the allowed number of inserted operators might 
be bounded because our proof of regularity does not apply when 
the order of perturbation becomes very high.

To extend the above discussion to the $n$-point 
functions at a later time, 
we only need to show the regularity of the $n$-point functions 
which are defined as the expectation values of the path ordered 
products of $\gz$ and ${^g\!\pi}$, for the Euclidean vacuum without 
the restriction that all the arguments are at the equal time, which 
will be a straight forward extension.  
In this manner, one can construct various excited states that  
are IR regular and free from the secular growth. 

\subsection{Comparison to the recent publications}

In the recent papers~\cite{SZ1210, ABG}, the absence of the secular
growth is also claimed. It would be
profitable to give a comparison between 
these works and our current work. 
First, in these papers, the item 1 raised
in Sec.~\ref{Sec:Intro}, {\it i.e.,} the presence of the canonical
system, which is equivalent to the original canonical system and whose
interaction Hamiltonian is composed only of the IR irrelevant operators,
is postulated, while this is not automatically guaranteed from the 
symmetry of the classical system. Second, in these papers, the mode
function in de Sitter spacetime, whose amplitude is given by a constant
Hubble parameter, is used in proving the conservation of the curvature
perturbation. This leads to the quantitative discrepancy in the
evaluation of the secular growth from ours. For instance, in Ref.~\cite{SZ1210}, the
locality of the solution $\dot{\tilde{\zeta}}^{(n)}_L(x, t)$ given in
Eq.~(22) of the paper is crucial in their proof. However, the locality
is not necessarily valid, once we take into account the fact that in the
chaotic inflation, the amplitude of the fluctuation becomes larger and larger in the distant
past as $\dot{\rho} \propto e^{- \int \dd \rho \varepsilon_1}$. When we
neglect this effect by setting 
$(\dot{\rho}/\sqrt{\varepsilon_1})^N$ in Eq.~(\ref{Exp:Ccond}) to
constant, the convergence condition is always satisfied (unless the
interaction coefficient $\hat{\lambda}$, composed of the horizon flow
functions, varies rapidly). Therefore our result does not contradict to the
conservation of the curvature perturbation that they claimed. 
The third point
is about the treatment of the UV contributions. In this paper, we have
not directly discussed about the UV renormalization. We simply assumed that the
UV divergent contributions, which are shown to be localized  
to the region where the two arguments of the Wightman functions are 
mutually almost light-like, 
can be renormalized by introducing the local counter terms. As long as the renormalization does not break
the dilatation symmetry of the classical action, our discussion can
hold. Recently, an interesting investigation about the UV
renormalization is pursued in Ref.~\cite{ABG}. 
It is claimed 
that a decaying composite operator in the free theory is kept
decaying also after the renormalization of loops. 
Although the non-trivial
assumptions such as the locality must be removed or verified, 
if this statement is correct, 
the conservation of the curvature perturbation can be shown also
in the presence of the loop corrections. We should, however, 
emphasize that
the conservation of the
curvature perturbation does not prohibit the appearance of the
logarithmic amplification, as we mentioned in Sec.~\ref{Sec:Intro}.

Finally, we also make a comment on the recent progress regarding the IR issues
of a test field in the exact de Sitter spacetime, which can be
interpreted as an approximation to the entropy mode. The
regularity of the loop corrections for the Euclidean vacuum is shown for
the massive scalar field by S.~Hollands~\cite{SH} and independently by D.~Marolf
and I.~Morrison~\cite{MM10, MMall, MM11}. By contrast, for a massless scalar field, the IR
regularity has not been shown and the absence of the secular growth 
is unclear~\cite{SH11, KK10, KK1012, KK11} (see also Ref.~\cite{Rajaraman:2010xd}). Although the adiabatic curvature
perturbation is a sort of massless field in the sense that the Wightman function $G^+(x,\, x')$
possesses the IR divergence and the long term correlation, the operation of the IR suppressing operators 
${\cal R}_x$, which appear by virtue of the residual \gauge
symmetry and by choosing the Euclidean vacuum, cures the singular
behaviour. Hence, it would be intriguing to discuss a massless 
field with the exact shift symmetry in the de Sitter spacetime, 
in comparison with the case of the adiabatic mode.

\acknowledgments
This work is supported by the Grant-in-Aid for the Global COE Program
"The Next Generation of Physics, Spun from Universality and Emergence"
from the Ministry of Education, Culture, Sports, Science and Technology
(MEXT) of Japan. T.~T. is supported by Monbukagakusho Grant-in-Aid for
Scientific Research Nos.~24103006, 24103001, 24111709, 21244033, 21111006.
Y.~U. is supported by the JSPS under Contact No.\ 21244033, MEC FPA under Contact No.\ 2007-66665-C02, and MICINN
project FPA under Contact No.\ 2009-20807-C02-02. We thank J.~Garriga,
M.~Sasaki, and R.~Woodard for their valuable comments. 
Y.~U. would like to B.~Allen and N.~Tsamis for the hospitality during the workshop
"Physics of de Sitter Spacetime" in the Max-Planck Institute for
Gravitational Physics.

\appendix

\section{Solving constraint equations}  \label{Sec:constraint}
In this section, we discuss about the boundary conditions of the
constraint equations, which are elliptic type. By expanding
the metric perturbations as 
$\tilde{\zeta}=\tilde{\zeta}_I + \tilde{\zeta}_2+ \cdots$,
$N=1+ \tilde{N}_1+ \tilde{N}_2+ \cdots$, and
$\tilde{N}_i=\tilde{N}_{i,1} + \tilde{N}_{i,2} + \cdots$,
the Hamiltonian constraint and the momentum constraints yield
\begin{align}
 & V \tilde{N}_n - 3 \dot{\rho} \dot{\tilde{\zeta}}_n + e^{-2\rho}
 \partial^2 \tilde{\zeta}_n + \dot{\rho} e^{-2\rho} \partial^i
 \tilde{N}_{i,n} =H_n\,, \label{HCn}\\
& 4 \partial_i \left( \dot{\rho}\tilde{N}_n - \dot{\tilde\zeta}_n \right) -
 e^{-2\rho} \partial^2 \tilde{N}_{i, n} + e^{-2\rho} \partial_i
 \partial^j \tilde{N}_{j, n} = M_{i, n}\,,
\end{align}
where $H_n$ and $M_{i,n}$ include $n$
interaction picture fields $\tilde{\zeta}_I$ in the combination
$\tilde{\zeta}_I-s$ or with differentiation. Eliminating $\tilde{N}_n$ from
these constraint equations, we obtain
\begin{align}
 & \left(1 - \frac{4\dot{\rho}^2}{V} \right) \partial_i \partial^j
 \tilde{N}_{j, n} - \partial^2 \tilde{N}_{i, n} + 2
 \frac{\dot{\phi}^2}{V} e^{2\rho} \partial_i \tilde{\zeta}_n 
 - 4 \frac{\dot{\rho}}{V} \partial_i \partial^2 \tilde{\zeta}
= C_{i,n}\,, \label{Cn}
\end{align} 
where we defined
\begin{align}
 & C_{i, n} :=  e^{2\rho} \left( M_{i, n} - \frac{4\dot{\rho}}{V} \partial_i H_n \right)\,.
\end{align}
Operating $\partial^i$ on Eq.~(\ref{Cn}), we obtain
\begin{align}
 & \partial^2 \partial^i \tilde{N}_{i, n} = \frac{\dot{\phi}^2}{2\dot{\rho}^2}
 e^{2\rho} \partial^2 \dot{\tilde{\zeta}}_n - \frac{1}{\dot{\rho}} \partial^4
 \tilde{\zeta}_n - \frac{V}{4\dot{\rho}^2} \partial^i C_{i,n} \,.
\end{align}
We solve this equation as follows,
\begin{align}
 & \partial^i \tilde{N}_{i, n}(x) = \frac{\dot{\phi}}{2\dot{\rho}^2}
 e^{2\rho} \dot{\tilde{\zeta}}_n(x) - \frac{1}{\dot{\rho}} \partial^2 \tilde{\zeta}_n(x)
 - \frac{V}{4\dot{\rho}^2} 
 \left[ \partial^{-2} \partial^i C_{i, n}(x) - G^L_n(x) \right]
\end{align}
where $G^L_n(x)$ is an arbitrary solution of the Laplace equation, 
{\it i.e.,} $\partial^2 G^L_n(x)=0$. Inserting this solution into
Eq.~(\ref{Cn}), we obtain 
\begin{align}
 & \partial^2\tilde{N}_{i, n} =  \partial_i \left[
 \frac{\dot{\phi}^2}{2\dot{\rho}^2} e^{2\rho}  \dot{\tilde{\zeta}}_n -
 \frac{1}{\dot{\rho}} \partial^2 \tilde{\zeta}_n - \frac{V}{4\dot{\rho}^2}
  \left( \partial^{-2} \partial^j C_{j, n} - G^L_n \right) 
  \right]   + 
   \partial_i\left( \partial^{-2} \partial^j C_{j, n} - G^L_n \right) 
  - C_{i, n}  \,.  \label{Eq:Cn2}
\end{align}
Again, introducing an arbitrary solution of the Laplace equation $G_{i,n}(x)$,
we solve Eq.~(\ref{Eq:Cn2}) as
\begin{align}
 \tilde{N}_{i, n}(x) &= \partial_i \partial^{-2} \left[
 \frac{\dot{\phi}^2}{2\dot{\rho}^2} e^{2\rho}  \dot{\tilde{\zeta}}_n(x) -
 \frac{1}{\dot{\rho}} \partial^2 \tilde{\zeta}_n(x) - \frac{V}{4\dot{\rho}^2}
  \left( \partial^{-2} \partial^j C_{j, n}(x) - G^L_n(x) \right)  
  \right] \cr
 & \qquad \quad +  \partial_i \partial^{-2}\left( \partial^{-2} \partial^j
 C_{j, n}(x) - G^L(x) \right) - \partial^{-2} C_{i, n}(x) +G_{i,n}(x)\,.
 \label{Exp:Nin}
\end{align}
Comparing the expression obtained by operating $\partial^i$ on
Eq.~(\ref{Exp:Nin}) with Eq.~(\ref{Eq:Cn2}), we obtain
\begin{align}
 & \partial^i G_{i, n} = \partial^i \partial^{-2} C_{i, n} 
  - \left( \partial^{-2} \partial^j  C_{j, n} - G^L \right)\,.
\end{align}
Using this expression, we rewrite the longitudinal part of $G_{i, n}$ as
\begin{align}
  G_{i, n} = \partial_i \partial^{-2} \left[ \partial^i \partial^{-2} C_{i, n} 
  - \left( \partial^{-2} \partial^j  C_{j, n} - G^L \right) \right]  + G_{i, n} -
 \partial_i \partial^{-2} \partial^j G_{j, n} \,. \label{Gin}
\end{align}
Inserting Eq.~(\ref{Gin}) into Eq.~(\ref{Exp:Nin}), we obtain
\begin{align}
  \tilde{N}_{i, n}(x) &= \partial_i \partial^{-2} \left[
 \frac{\dot{\phi}^2}{2\dot{\rho}^2} e^{2\rho}  \dot{\tilde{\zeta}}_n(x) -
 \frac{1}{\dot{\rho}} \partial^2 \tilde{\zeta}_n(x) - \frac{V}{4\dot{\rho}^2}
  \left( \partial^{-2} \partial^j C_{j, n}(x) - G^L_n(x) \right)  
  \right] \cr
 & \qquad \quad - \left( \delta_i\!^j - \partial_i \partial^{-2}
 \partial^j  \right) \left(  \partial^{-2} C_{j, n}(x) - G_{j, n}(x)
 \right)\,. \label{Exp:Nin2}
\end{align}

When we perform quantization in the whole universe, it is natural to
request the regularity of the perturbation at the spatial infinity. This
requirement uniquely fixes $G^L_n$ and the transverse part of 
$G_{i,n}$. Then, the shift vector depends on the curvature perturbation
$\tilde{\zeta}$ of the whole universe. To show the IR regularity, here
we employ another boundary condition which requests that the integration
region of the inverse Laplacian $\partial^{-2}$ is confined to around
the local observable region ${\cal O}$. As is shown in Refs.~\cite{IRgauge_L, IRgauge}, the
degrees of freedom in changing the boundary condition can be understood
as the \gauge degrees of freedom in the local universe. Therefore,
the operator $\gR$ is invariant under the change of the boundary
condition.

Adjusting the solutions of the Laplace equations $G^L_n(x)$ and
$G_{i,n}(x)$, we can change the boundary condition for
$\partial^{-2}$ so that the integration region is limited. We fix the function $G^L_n(x)$, requesting 
\begin{align}
 & \partial^{-2} W_{L_t}(\bm{x}) \partial^j C_{j, n}(x)  =  \partial^{-2}
 \partial^j C_{j, n}(x) - G^L_n(x)  \,,
\end{align}
where we inserted the window function which takes a non-vanishing value
only within the vicinity of the observable region ${\cal O}$. If we evaluate the term in the first line of
Eq.~(\ref{Exp:Nin2}) by using the Laplacian inverse with two different boundary conditions,
$\partial^{-2}_1$ and $\partial^{-2}_2$, the difference is given by
\begin{align}
 & \partial^2 \left(  \partial_i \partial^{-2}_1 \left[ \cdots \right] -
 \partial_i \partial^{-2}_2 \left[ \cdots \right]  \right) =0\,, \qquad
 \qquad  \partial^i \left(  \partial_i \partial^{-2}_1 \left[ \cdots \right] -
 \partial_i \partial^{-2}_2 \left[ \cdots \right]  \right) =0\,,
\end{align}
where we abbreviated the terms in the square bracket. Therefore, the
change of the boundary condition for the Laplacian inverse
can be absorbed by the transverse mode of $G_{i, n}(x)$. Fixing the boundary condition of
$\partial^{-2}$ so that the integration region is restricted to the
vicinity of the observable region, we obtain
\begin{align}
  \tilde{N}_{i, n}(x) &= \partial_i \partial^{-2} W_{L_t}(\bm{x}) \left[
 \frac{\dot{\phi}^2}{2\dot{\rho}^2} e^{2\rho}  \dot{\tilde{\zeta}}_n(x) -
 \frac{1}{\dot{\rho}} \partial^2 \tilde{\zeta}_n(x) - \frac{V}{4\dot{\rho}^2}
   \partial^{-2} W_{L_t}(\bm{x}) \partial^j C_{j, n}(x) 
  \right] \cr
 & \qquad \quad - \left( \delta_i\!^j - \partial_i \partial^{-2}
 \partial^j  \right)  \partial^{-2}  W_{L_t}(\bm{x}) C_{j, n}(x)\,. \label{Exp:Nin3}
\end{align}
Inserting this solution into Eq.~(\ref{HCn}), we can also obtain the
lapse function whose support of the Laplacian inverse $\partial^{-2}$ is also
confined.

\end{document}